\documentclass[aps,epsfig,showpacs]{revtex4}
\usepackage{amsmath}
\usepackage{amssymb}
\usepackage{psfrag}
\usepackage{epsfig}
\usepackage{latexsym}
\newcommand{\be}{\begin{eqnarray}}
\newcommand{\ee}{\end{eqnarray}}
\newcommand{\la}{\langle}
\newcommand{\ra}{\rangle}

\begin{document}
\title{Is Econophysics a Solid  Science?}
\thanks{This work has been commissioned by 
the Editor of {\it Acta Physica Polonica B}. It has been financed by 
Stowarzyszenie Zbiorowego Zarz¹dzania Prawami Autorskimi Twórców 
Dzie³ Naukowych i Technicznych KOPIPOL z siedzib¹ w Kielcach, 
from the income coming from implementation of Art.~20 of the 
law on authorship and related to its regulations.}

\author{Z. Burda, J. Jurkiewicz, M.A. Nowak}

\affiliation{
\vspace{0.4cm}
$^2$ Institute of Physics, Jagellonian University,
ul. Reymonta 4, 30-059~Krak\'ow, Poland. \\
}

\begin{abstract}
\noindent
Econophysics is an approach to quantitative economy
using ideas, models, conceptual and computational methods
of statistical physics. In recent years many of physical
theories like theory of turbulence, scaling,
random matrix theory or renormalization group were successfully
applied to economy giving a boost to
modern computational techniques of data analysis, risk management,
artificial markets, macro-economy, {\it etc.} Econophysics became
a regular discipline covering a large spectrum of
 problems of modern economy. It is impossible
to review the whole field in a short paper.
Here we shall instead attempt to give a
flavor of how econophysics approaches economical
problems by discussing one particular issue
as an example: the emergence and consequences
of large scale regularities, which in particular
occur in the presence of fat tails in
probability distributions in
macro-economy and quantitative finance.
\end{abstract}

\maketitle

\section{Introduction}
Half a decade ago, a word ``econophysics'' started to circulate
in the community of physicists. In July 1997, ``Workshop on Econophysics''
was organized in Budapest by Imre Kondor and
Janos Kertesz~\cite{BUDAPEST}.

Followed by several other dedicated meetings, the field matured,
reaching the state when textbooks on the
subject, written by the pioneers in the field,
started to appear~\cite{STANMAN,BOUPOT,ROEHNER}.

The name ``econophysics'',  a hybrid of ``economy'' and ``physics'',
was coined to describe applications of methods of statistical physics
to economy in general. In practice, majority
of the research concerned the  finances.
In such a way, physicists entered officially and scientifically
the field of financial engineering. On top of similar statistical
methods used by financial mathematicians  (although formulated in not so
formal or ``high-brow'' fashion as in the textbooks on financial mathematics),
physicists concentrated on the analysis of experimental data
using tools borrowed from the analysis of real complex systems.

Commissioned  by the Editorial Board of {\it Acta Physica Polonica B} to pre\-sent an
overview of the ``econophysics'' oriented towards  a physicist
 who never really entered this interdisciplinary area,
we faced the danger of an attempt to present the status of the discipline
which is still {\it  in statu nascendi}, reviewed by authors
biased strongly
by their personal views related to their (limited) own
research in the newborn field.
Therefore this mini-review is to a large extent
a collection of thoughts and results
from works of the three authors.
As such, it is not intended to cover the whole field
which has become a large discipline with many
sub-branches by now but instead
to present a modest sampler of scientific methods borrowed
from physics  to describe economical ``data''.
We restricted to the methods which were natural extrapolation of
those used in our own research in fundamental science (quantum gravity,
random matrices, random geometry, complex systems).
As a guiding line through this mini-review
we have chosen  power laws due to their omni-presence
in economical data.

The review is organized as follows. We  begin with
a historical introduction arguing that
despite the name ``econophysics'' entered the scientific language
only half a decade ago,
connections and interplay between physics and economy are more than
hundred years old. The official  marriage of disciplines of
economy, often understood as an art, and physics
being an example of a hard science,
has been preceded by the continuous development of scientific
methodology for a long time. One could even say that
the official recognition of the close links came surprisingly late.

In the second part we concentrate on  power-laws
in economy. Using the system size criterion we
divide the economical world into
macro-, meso- and microscopic objects:
the first of which are related to macro-economy,
the second to stock markets
and the third to individual companies.
The levels are intertwined. In macro-economy one
observes fat tails in the wealth and income distributions.
Analysis of stock markets clearly shows the presence of
large scale events, which can be described by  probability
distributions with fat tails. The same concerns price fluctuations
of individual companies. At each of these regimes,
one uses slightly different tools of the analysis.
As we shall argue they all have common roots
in the theory of large numbers. We shall start with the macro-economical
application where we discuss the wealth and income distributions.
Then we switch to the micro- and mesoscopic regimes where
we shall concentrate on statistical properties of
the system of fluctuating assets
and on a question how the signal can be extracted
in such a system. The natural language for the description
of such a system is provided by the random matrix theory.
We shall discuss the central limit theorem
for random matrices and its consequences.

In the last, third part we very briefly mention other
active areas of research which have recently
attracted attention of the econophysics community.
We also try to speculate on potential dangers of the approach,
which may arise if methods of physics are adapted to economy
to blindly. We believe that the success of scientific
methods for economic applications requires
broader scientific methodology, borrowing largely not only from
physics, but also from other domains of science,
mainly the theory of adaptive systems,
studies  of computer networks  or the  analysis of complex systems.
Only successful  evolution  of ``econophysics''  into
``econoscience'', accompanied by rigid constraints
based on careful analysis of empirical data,
gives economy a chance  to become a predictive
theory at a high confidence level,
 and may acquire a status of  a ``hard science''.
We conclude that achievement of this goal,
although not easy, is certainly possible.

\section{Historical background}

At a first glance, economy and physics do not seem to be related.
Despite the fact that the literature is full of examples of
famous physicists being interested in
economic or financial problems, these examples are usually  treated
as adventures, and are sometimes anecdotical.
Some well known cases are:
\begin{itemize}
\vspace{-2mm}
\item unsuccessful predictions of stock prices
by sir Isaac Newton, and in consequence, his  terrible loss in 1720
 of 20000 pounds
  in South Sea speculation bubble~\cite{CHANCELLOR},\vspace{-2mm}
\item successful  management  of the fund for widows
   of Goetingen professors, performed by Carl Friedrich Gauss,\vspace{-2mm}
\item explanation of the Brownian random walk and the formulation of
   the Chapman-Kolmogorov condition for Markovian processes
   by Louis Bachelier in his PhD thesis on the theory of speculation
   done 5 years before the Smoluchowski's and Einstein's works
   on diffusion, on the basis of the observations of
   price movements on Paris stock-market~\cite{BACHELIER}
\end{itemize}
and few others.
These examples put forward the thesis which may sound revolutionary for a
contemporary econophysicist:
It was the  economy  which followed physics,
and not {\it vice versa} --- studies of the XVIII and XIX century classical
physics made a dramatic
impact on economy, and the work  was done mostly by the economists,
who tried to follow the
scientific methodology of physical sciences~(see {\em eg}~\cite{DELISO,MIROWSKI}).

As a first example we mention the father of classical  economy, Adam Smith.
In his work {\it ``The principles which lead and direct philosophical
enquires: illustrated by the history of astronomy''}, Smith exemplifies
the methodology of science by stressing the role of observing the regularities
and then constructing theories (called by Smith ``imaginary machines'')
reproducing the observations. Using the astronomy as a reference point
was not accidental --- it was the celestial mechanics, and the impressive
amount of astronomical data, which dominated science in several cultures.
It is rather amazing, that this analysis was done by a person,
who is primarily identified as an economist, and not as a ``physical scientist''.
In the end of XVIII and  in XIX century, Newton's theories
were transformed into more modern language
of  analytical mechanics in the works  of Lagrange, Hamilton and others
(actually, this is the formulation still
used in textbooks of mechanics today).
The beauty and power of the analytical mechanism did not escape the attention
of the economists. In particular, the concepts of
mechanics were  considered as an ideal
tool to be used in mathematization of economy.
Again, it is perhaps surprising for a contemporary financial engineer
that mathematics entered economy through physics!
Economists like Walras, Jevons, Fisher, Pareto tried to map the formalism
of physics onto the formalism of economy, replacing material points by
economic agents, finding the analogy of the potential energy
 represented by ``utility'', and then evolving the systems by the analogs
of principle of minimal action~\cite{MIROWSKI}.
 That fascination with mechanics
went so far, that economists  were even  building   mechanical models
 illustrating the concept of economical equilibrium.
The enchantment with classical physics dated till the first half of the
XX century. Again, it is surprising for a physicist, that the
conceptual
revolution done by Boltzmann (concepts of probability) and quantum
mechanics (another meaning of probability), were missed for so long by the
economists.
Visionary suggestions by  Majorana~\cite{MAJORANA} in the 30's
to use statistical physics in social science were at that time
not explored neither by physicists nor by economists.

It is surprising even more, if we recall the example
of the already mentioned
Louis Bachelier, who formulated the theory of Brownian motion
on the basis of economic data and moreover 5 years before the seminal
works by Einstein and Smoluchowski. Almost half a century after the
defense of his thesis {\it ``Jeu de speculation''} (not appreciated very much
by his advisor, Henri Poincaré), the ideas of Bachelier were discovered
 in the economy departments of major American universities.
A slight modification of the Bachelier stochastic process (basically,
changing the additive noise into the multiplicative) lead Osborne
and Samuelson\cite{OSBORNE}  to the fundamental stochastic equation
governing the evolution of stock prices and serves as a cornerstone of the
famous
theory of Black, Scholes and Merton for calculating the correct price
of an option. Technically, the Black--Scholes formula is just the solution of
the heat equation, with a peculiar boundary condition.
The incredible practical success of option-pricing formulae perhaps
lured economists and financial engineers a bit, and maybe, to some extent,
was responsible for the spectacular crash on Wall Street in
August and September 1998 which ricocheted over the other markets.

Taking into account several discoveries done in  physics, one could say
that perhaps in the 80' the economists  missed a lesson from physics.
Concepts of a random walk were formulated using the assumption of
the Gaussian character of a stochastic process.
As such, the movement of prices was considered as memoryless,
with almost negligible effects of large deviations, exponentially
screened in the Gaussian world. Actually already in the 60' Mandelbrot
pointed certain selfsimilarity of the behavior of commodities
(cotton prices) over different time scales, interpreted as the
appearance of power law. Today, for a physicist,
familiar with critical phenomena, the concept of a power law
and large fluctuations is rather obvious, although she or he may not be
familiar with the fact
that the main concepts of fractal behavior, spelled by Mandelbrot in
70', were predecessed by his  study of cotton prices, done a decade earlier.
Actually, stock markets exhibited
 large fluctuations (power behavior is usually
named as ``fat'' or ``heavy'' tail
 behavior), but rather a limited interest
in this behavior in the 90' was caused to large extent
by the reservation of financial mathematics,
 lacking  powerful mathematical methods
(like Ito calculus) suited for processes with divergent moments.

The second major factor, changing the Gaussian world was a computer.
In the last 40 years the performance of the computers
had increased by six orders of magnitude.
This fact had to have a crucial impact on economy.
First, the speed and the range of transactions had changed drastically.
In such a way computer started involuntarily to serve as an amplifier
of fluctuations. Second, the economies and markets  started to watch
each other more closely, since computer possibilities allowed for
collecting exponentially more data.

In this way, several nontrivial couplings started to appear in
economical systems, leading to nonlinearities.
Nonlinear behavior and overestimation of the Gaussian principle
for fluctuations were responsible for the Black Monday Crash in 1987,
and the crisis in August and September 1998.

That shock had however also a positive impact visualizing the
importance of the non-linear effects.
Already Poincaré has pointed the possibility of unpredictability in
a {\it nonlinear} dynamical system,
establishing the foundations of  the chaotic behavior.
The study of chaos turned out to be a major branch of theoretical physics.
It was only a question of time, how fast these ideas will start to appear
in economy. Ironically, Poincaré, who did not appreciate
Bachelier's results, made himself a  large impact on real
complex systems as one of the discoverers of chaotic behavior
in dynamical systems.
Nowadays studies of chaos,
 self-organized criticality, cellular automata
and neural networks are seriously taken into account as
economical and financial tools.

One of the benefits of the computers was that
economic systems started to {\it save}   more and more data.
Today markets collect incredible amount of data (practically
they remember every transaction). This
triggers the need for new methodologies,
able to manage the data.
In particular, the
data started to be analyzed using methods,
borrowed widely from physics,
where seeking for regularities and for
unconventional
correlations is mandatory.

It was perhaps the reason, why several institutions (however, more
financial than
devoted to study the problems of macroeconomy) started hiring physicists
as their ``quants'' or ``rocket scientists''. In the last ten years,
another tendency appeared --- physicists started to study economy
scientifically. Several educational or research institutions devoted to
study  complexity launched the research programs in economy and
financial engineering. These studies were devoted mostly to
quantitative finance.
To a large extent, it was triggered by vast amount of data accessible
in this field. In such a way, physics started to play the role of
financial mathematics --- sometimes rephrasing the mathematical
constructions in
the language of physics,    sometimes applying methods developed solely
in physics, usually at the level of various effective theories of complex
systems.
Name ``econophysics'', often attributed to
the activity of physicists in this field, is in our opinion rather misleading
--- perhaps ``the  physics of finances''
is more adequate or even ``statistical phynance'' as  J.P. Bouchaud jokes.
Moreover, as we speculate in the conclusions of this work,
name ``physics'' may be to
restrictive to include majority of  the tools of financial analysis.

Probably the most challenging questions in economy are those related to
macro-economy. Extrapolating the historical perspective,
briefly sketched above, to the future, one can expect
methods of physics, especially those used in
studies of complex and nonlinear systems,
to make an impact on this field in the nearest future.
In this case the meaning of econophysics would be similar to
``physical economy'', and econophysics  could  be viewed
as a physicists'  realization  of XIX century economists' dream.

\section{Macro-economy}

Let us now turn to an example of econophysical
reasoning in macro-economy.
The term macro-economy has in general a double meaning:
of a science which deals with large scale phenomena in economical
systems and of a system which is the subject
of the macro-economical studies.
Such a macro-economical system is a complex system which consists of
many individuals interacting with each other.
The individuals function in the background provided by the
legal and institutional frames. Individuals differ in abilities,
education, mentality, historical and cultural background {\it etc}.
They enter the system with different financial and cultural
initial conditions. Each of them has his own vision of what
is important and of what she or he is willing and able to achieve.
It is clear that one cannot formulate
a general theory of needs and financial possibilities of a single
individual or to create an economical profile of a typical member
of such a complicated system.  There are too many random factors
to be taken into account. They change in time: sometimes
slowly, sometimes faster, sometimes abruptly and
in an unpredictable way.
Every day some individuals leave the system, some new enter it.
It is impossible to follow individual changes.
One can however control their statistics. Actually, it is
the statistics which shapes the system on large macro-economical
scale and drives the large scale phenomena observed
in the whole macro-system.

The aim of macro-economical studies is to extract
important factors, understand their mutual relations
and describe the development of past events.
The ultimate goal is to reach
a level of understanding which would also permit to predict
the reaction of the system to the change of macro-economical
parameters in the future.
Having such a knowledge at hand, macro-economists would
be able to stimulate the optimal evolution by appropriately
adjusting the macro-economical parameters.
This level of understanding goes far beyond a formal
description and requires modeling and understanding of
fundamental principles which are difficult because
of the complexity of the problem. Clearly, a model whose main
ambition would be to realistically take into account all parameters
and factors characterizing the whole network of dependencies
in such a complex system would fail to be comprehensive
and solvable.  One would not be able
to learn anything from such a model. It would be even to
complicated to properly reflect what it actually intends to describe.

Obviously, one has to find a way of simplifying the underlying
complexity to the level which enables a formulation of a treatable model.
A danger of a simplification of a complex and non-linear problem
is that by a tiny modification one can loose an important
part of the information or introduce some artificial effects.
There are two possible approaches to the problem of modeling
complexity.  One way is to follow a phenomenological
reduction scheme. The first step is to introduce
effective phenomenological quantities which encode the most
important part of the reduced information.
Of course, it is very difficult to quantify many important
factors like cultural potential, historical background or
influence of a change in particular law which for example
regulates relation between employers and employees {\it etc.}
Such factors play crucial role in the outcoming shape of the
macro-system. The next step is to determine mutual
dependencies of these quantities. This procedure usually
leads to a set of non-linear differential equations describing
evolution of the phenomenological quantities as a function
of other parameters. At this level a new complication occurs.
It is well known that nonlinear equations generally possess
a very complicated spectrum of solutions whose stability depends
on precise values of the parameters. Sometimes tiny changes of
parameters which are irrelevant, from the point of view of
the macro-economy, may be significant for
the underlying mathematics, and opposite. In other words,
a formal mathematical solution
does not always carry a realistic economical
information. One has to distinguish between the real and
artificial effects. It is not always easy and one should
be aware of limitations steaming from the
complexity and non-linearity.

An alternative approach is the search for universal laws which govern
the behavior of the complex system. Such laws may uncover
global regularities which are insensitive to tiny changes
of parameters within a given class of parameters.
Such laws also provide a classification of possible universal
large scale behaviors which can occur in the system and which can be
used as a first order approximation in the course of gaining
insight into the mechanisms driving the system.

This approach has been successfully used in theoretical
physics for a long time where for a given model
one is able
with the aid of the renormalization group ideas to determine
so called fixed points, each of which
being related to one universality class of the model \cite{kw}.
The space of all possible classes of different large scale
behaviors of the model is divided into subspaces called
domains (or basins) of attraction of those fixed points.
The universal properties of any theory
within a domain of attraction of a given fixed point
are entirely determined by the properties of the renormalization
group map in the nearest neighborhood of the fixed point.
The number of domains of attraction is usually small. Thus
typically one has only a few distinct universal large scale
behaviors despite the original theory has infinitely many
degrees of freedom and infinitely many coupling constants
controlling the mutual interactions of those degrees of freedom.
Macro-economical systems are
in this respect very similar to field theoretical ones.

Another well known example of the emergence of universal laws
is the central limit theorem. Saying not rigorously,
the central limit theorem tells us that the
sum of many independent identically
distributed random numbers
polled from a distribution with a finite average and a
finite variance obeys a Gaussian law
with the mean and the variance which scale
with the number of terms in the sum
independently of the particular shape of the distribution.
One could say that all distributions with finite variance
belong to the Gaussian basin of attraction. The Gaussian
distribution is stable. Stable distributions play here the
role of fixed points. We see that
a regularity emerges for large sums telling us that
all details of the original distribution except the mean and the
variance get forgotten in the course of enlarging the number
of terms in the sum. Distributions with infinite variance belong
to the Lévy universality classes (or saying equivalently
to the basin of attraction of the Lévy distributions) \cite{f,gk}.

One expects the large scale phenomena
in economy to display a universal character because
they result from a large number of events which are
driven by laws of the same system and which contribute
to the same statistics.

In this paper we shall take the latter approach.
We shall be looking for general laws which
describe large scale behavior of economical systems.
We shall try to deduce them from
assumptions as simple as possible,
which define  certain universality classes.
Small refinements and perturbations are believed not
to change the universality
class of the large scale behavior.
As an example, in the next section we shall concentrate
on the issue of the wealth and income distribution.
This issue, addressed already by Adam Smith, still
 stands in the central place
in the macro-economical research.

\section{Wealth and income distributions}

As mentioned above, we  argue that the laws governing
distributions can be deduced from the mathematics of large numbers.
A simple assumption about the nature of wealth fluctuations
seems to capture properly the microscopical mechanism
which in the large scale leads to the emergence of
laws known for a long time from empirical studies  in macro-economy.
The first law, discovered
by Pareto more than one hundred years ago \cite{p}, tells us that
the wealth distribution of the richest part of the society
is controlled by the power-law tail
\begin{equation}
d w \ p(w) \sim
\frac{\alpha A^\alpha d w}{w^{1+\alpha}} \quad
\mbox {for} \quad w \gg w_0\,.
\label{pareto}
\end{equation}
Here $p(w) dw$ stands for a probability that a randomly
chosen member of the macro-economical system possesses the wealth
between $w$ and $w+dw$; $w_0$ has the meaning
of a typical value of the individual's wealth in this system.
The exponent $\alpha$ is called the Pareto index.
Pareto himself suspected that there may exist an
underlying mechanism which singles out a particular
fixed value of this index. Today we know that it is not
true. The value of the Pareto index $\alpha$
changes from macro-economy to macro-economy \cite{s}.
It also varies in time. The empirical estimates show that
a value
of the Pareto index in real macro-economical systems
fluctuates around two.

It is worth discussing the consequences of the presence of
the power-law tail in the probability distribution.
An immediate consequence is that the probability that
a random person from the richer part of the society
is $\lambda$ times richer than another person with
wealth $w$
\begin{equation}
\frac{p(\lambda w)}{p(w)} \sim \lambda^{1+\alpha}
\label{scaling}
\end{equation}
is independent of $w$. This distribution is scale-free,
reflecting a certain self-similarity of the structure
of the richest class. Actually
the scale appears in the problem
through the parameter $w_0$ which provides the lower
cut-off above which $w \gg w_0$ the power-law part
of the distribution sets in. The scale is provided by
prices of elementary goods which one needs
to function in the system, like for instance
prices of houses, cars, {\it etc}. Being rich means
to be far above this scale, to the degree that
it does not matter how much the basic things cost.

Let us take a closer look at some values to gain the
intuition about the consequences of the Pareto.
For $\lambda=10$ and $\alpha=2$,
the factor on the right hand side of (\ref{scaling})
is $10^{-3}$. Thus for $\alpha=2$
the Pareto law predicts that the number of people ten times
richer is roughly one thousand times smaller.
The suppression factor is very sensitive to $\alpha$.
If the value of $\alpha$ moves towards unity,
the suppression factor decreases, and for $\lambda=10$
it is only $10^{-2}$.
In other words, in the macro-economy with a smaller value of $\alpha$
the tail of the distribution is fatter. This
leaves more space for rich individuals.
Thus one intuitively expects that for smaller $\alpha$
the macro-economy is more liberal. In a more restrictive
macro-economical system the Pareto exponent $\alpha$ is larger
and hence the richer population is suppressed.

The presence of heavy tails in empirical data
is relatively easy to detect. One just observes cases lying
far beyond the range suggested by standard estimators of
the mean and width of the distribution. What is however
difficult is to quantitatively estimate the values of
the Pareto index. The reason for this is actually very simple. As
follows from the discussion above,
cases with a very large deviation from the mean are relatively
rare --- much more rare than those in the bulk of the distribution.
Thus the statistics in the tail is very poor. The effect
of small statistics is additionally amplified by the fact
that for a given macro-economical
system one can carry only one measurement of the wealth distribution.
One thus has  only one statistically independent
sample.
Secondly, the crossover between the bulk of the distribution
coming from the lower and middle classes and the tail coming from
the richest is smeared and therefore it is not entirely clear where
the Pareto law sets in: the position
of the termination point of the Pareto tail is not unique.
This uncertainty introduces a bias to the estimators.

Moreover, gathering data about personal wealth
and income is a delicate matter.
It is technically very difficult, close to impossible,
to collect the unbiased data, which would be free of
personal, social or political factors.

Here we shall discuss only the difficulty related
to poor statistics. Having the wealth distribution $p(w) d w$
one can easily estimate the probability that the wealth
of a random member of the macro-economy exceeds a certain value
$W$
\begin{equation}
P(W) = \int\limits_W^\infty  d w \ p(w) \,.
\label{cumul}
\end{equation}
For the particular form of the power law (\ref{pareto}) this
probability can be calculated to be
\begin{equation}
P(W) \sim \left(\frac{A}{W}\right)^\alpha \quad \mbox{for} \quad W\gg
w_0\,.
\end{equation}
In the population of $N$ people the number of individuals
whose wealth exceeds $W$ is roughly of the order $P(W) N$. Thus
denoting the wealth of the richest by $W_{\rm max}$,
one can estimate $P(W_{\rm max}) N \approx 1$  and hence
\begin{equation}
W_{\rm max} \sim AN^{1/\alpha}\,.
\end{equation}
A more involved analysis allows one to determine the
distribution of wealth of the richest
in the macro-economy with the power-law tail
to be given by the Fréchet distribution \cite{mxs}
\begin{equation}
d \omega \ p_F(\omega) = d \omega \,
\frac{\alpha}{\omega^{1+\alpha}} e^{-\omega^{-\alpha}}
= d e^{-\omega^{-\alpha}}\,,
\label{pf}
\end{equation}
where $\omega$ is a rescaled variable $\omega = W_{\rm max}/AN^{1/\alpha}$.
The distribution of the maximal wealth
inherits thus the power-law tail from the
original wealth distribution $p(w) d w$. This means that
in some realizations of the same macro-system
the richest may be much richer that the richest in other
realizations. As a consequence, the maximal wealth may undergo
strong fluctuations and so may the whole empirical data points
in the Pareto tail.
This is an additional factor which makes
the quantitative analysis
of the Pareto tail in the macro-economical data difficult.

\begin{figure}[htb]
\centerline{\epsfig{file=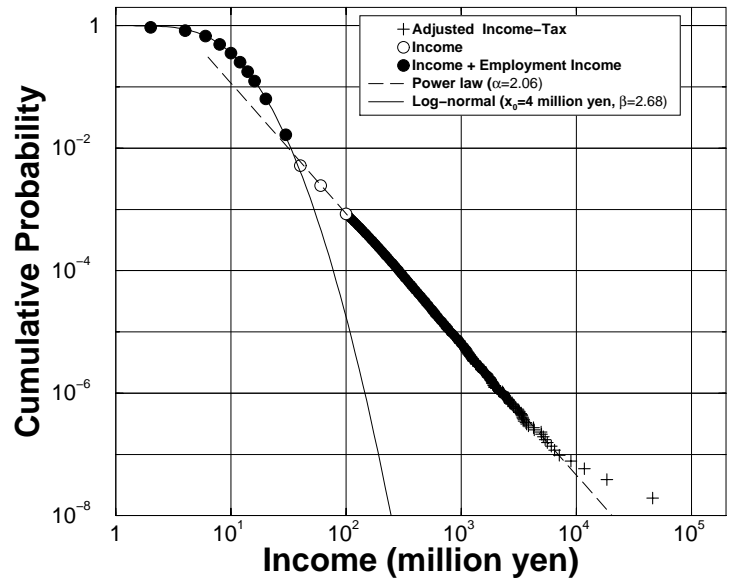,width=9.5cm}}
\caption{The power law and the lognormal fits to the 1998 Japanese
income data.
The solid line represents the lognormal fit with $x_0=4$
million yen and $\beta=2.68$. The straight dashed line represents
the power law fit with $\alpha=2.06$.
Reprinted from the paper \cite{s}
with the kind permission of the author.
The data sets presented in the figure come from
three different sources. The corresponding
data points are denoted by different symbols in the figure.
See \cite{s} for the detailed description.}
\label{soumaf}
\end{figure}

It is much easier to study empirically the distribution
in the range of smaller wealths. The
statistics is much better in this case
since the poor and middle class sectors
are more numerous. Also the income declarations are
statistically more reliable.
In  effect, the flow of wealth is much easier to control.
The statistics is thus less biased. Surprisingly the empirical
law which governs this part of the
income and wealth distributions
was discovered only four decades after the Pareto
law. It was discovered by Gibrat and named after him \cite{g}.
According to this law the wealth and income distributions
for the lower and middle classes obey the log-normal law
\begin{equation}
d w \, p(w) =
\frac{d w}{w}
\frac{1}{\sqrt{2\pi\sigma^2}}
\exp - \frac{\log^2 w/w_0}{2\sigma^2}\,.
\label{gibrat}
\end{equation}
The cumulative
probability $P(W)$ that the wealth of a random member
of the Gibrat macro-economy exceeds $W$ is given by
\begin{equation}
P(W) = \int\limits_W^\infty  d w \, p(w) =
\frac{1}{2} \mbox{erfc}\left( \frac{\log w/w_0}{\sqrt{2} \sigma}
\right)\,.
\end{equation}
All moments of the Gibrat distribution
are finite $\langle w^k \rangle = w_0^n \exp \sigma^2 n^2/2$.
The parameter $\sigma^2$ gives a typical width of
fluctuations of the order of magnitude of $w$ around $w_0$.
The values $w$ which deviate from $w_0$ by few $\sigma$
are strongly suppressed for the Gibrat distribution.
Sometimes to distinguish between the Gibrat and Pareto
distributions for large $W$ one draws the cumulative distributions
in the log-log plot \cite{s}. The plot $\log P(W)$ versus $\log W$
has a parabolic shape for the Gibrat distribution
when $W$ goes to infinity,
while the corresponding plot for the Pareto distribution
is a straight line (see Fig.~\ref{soumaf}),
This makes an enormous difference between
the Pareto and Gibrat laws in the range of large wealths.

Let us discuss mathematical mechanisms which may underlie
the Gibrat and Pareto laws. Imagine a random individual
in the system. Denote her or his wealth at a time $t$ by $w_t$,
and by $w_{t+1}$ at a later time, separated by
one unit $\varepsilon$ of time. The wealth could increase
or decrease by some factor $\lambda_t$ \cite{OSBORNE}
\begin{equation}
w_{t+1} = \lambda_{t+1} w_t\,.
\label{mult}
\end{equation}
In general this factor may itself depend on many factors like which
particular individual we picked up to look at,
with whom she or he interacts in the system,
what is his or her current financial situation {\it etc}.
In the simplest approximation, which would be called in
physics a mean-field approximation,
we assume this factor to be a random number from the
representative distribution which statistically
characterizes the whole system. Further,
the distribution is assumed to
depend neither on time nor on the current wealth.
The first assumption means that the process is stationary,
and the second that it is linear in wealth. Although all
this seems to be a crude approximation,
the essential point is that it may be enough to
capture the general properties of the related universality class.
What seems to be significant in the assumption is that
the variation of the wealth is described
by a multiplicative rather than
an additive process. Hopefully the large scale behavior
which we want to deduce from this assumption is
representative for a larger class including also
more complex processes.

The assumed multiplicative nature of changes seems to well
reflect the economical reality in which the primary objects which
fluctuate are the rates of exchange understood in a broad sense:
rates between goods, currencies, money, real estate {\it etc.}
The prices of stocks also belong to this category. The change
of wealth is proportional to the change of the exchange rate
which implies the multiplicative nature of changes. In a diversified
portfolios the situation is a little more complicated as
we shall discuss later.

It is convenient to parameterize the changes of the factor scale
$\lambda_t$ by the quantity $r_t$ which is related to
$\lambda_t$ as follows: $\lambda_t = \exp r_t $ or
equivalently as
\begin{equation}
r_t = \log \lambda_t = \log w_{t+1} - \log w_t\,.
\label{return}
\end{equation}
When the time
unit $\varepsilon$ between $t$ and $t+1$ is small, the
factor $\lambda_t$ is close to unity. In this case it can
be substituted by $\lambda_t = 1 + r_t + \dots$ which
gives the meaning of an instantaneous return
to the quantity $r_t$. The parameterization
$\lambda_t = \exp r_t$ automatically takes care
of the positive definiteness of the
scale factor $\lambda_t$: for $r_t$ fluctuating in the range
$(-\infty,+\infty)$, $\lambda_t$ fluctuates in the range $(0,+\infty)$.
In the simplest model
the statistical information about the returns $r_t$ is encoded
in a probability distribution $\rho_\varepsilon(r) d r$
which  characterizes  the system. Successive returns
$r_t$ are assumed to be random numbers polled
from the same distribution $\rho_\varepsilon(r)$.
The wealth $w_T$ and the return $R_T$
after the time $\tau = T\varepsilon$
which elapsed from the  moment $t=0$, is
given by the equation
\begin{equation}
R_T = \log \frac{w_T}{w_0} = \sum_{t=1}^T r_t
\label{add}
\end{equation}
as can be directly deduced from the equation (\ref{mult}).
If the mean and the variance
\begin{eqnarray}
\bar{r} &=& \langle r \rangle_\varepsilon\,,  \nonumber\\
\sigma^2 &=& \langle (r-\bar{r})^2\rangle_\varepsilon
\label{mv}
\end{eqnarray}
of the distribution $\rho_\varepsilon(r)$ are finite, the
distribution of the return $R_T$ approaches the normal
law with the density
\begin{equation}
d R_T \ P_T(R_T) =
d R_T \ \frac{1}{\sqrt{2\pi T{\sigma}^2}}
\exp -\frac{(R_T - T \bar{r})^2}{2 T{\sigma}^2}
\label{prob}
\end{equation}
as follows from the central limit theorem. We use
the relation between the return $R_T$ and the wealth
$w_T$ (\ref{add}) to obtain the distribution of wealth
\begin{equation}
d w_T \ p_T(w_T) =
\frac{ d w_T }{w_T}
\frac{1}{\sqrt{2\pi T{\sigma}^2}}
\exp -\frac{\log^2 w_T/w_0 e^{\bar{r} T} }{2 T{\sigma}^2}\,.
\label{ln}
\end{equation}
This is the Gibrat law \cite{g}.
The typical wealth of individuals in the system
changes in time as $w_0 e^{\bar{r} T}$ and
the range of the order of magnitude of fluctuations
as $\sqrt{T}{\sigma}$. A few comments are in order.
A typical wealth of the system increases in time
if the return $\bar{r}$ is positive and decreases if the
return is negative. It is constant for $\bar{r}=0$.
If one assumes it changes slowly (adiabatically) in time one
can think of $R$ as a sort of an averaged return. Thus in
some periods the total wealth may grow and in some diminish.

The width of the wealth fluctuations
which is given in the formula (\ref{ln}) by
$2T{\sigma}^2$, grows in the model even if one assumes
adiabatic changes: $\int^T d t \ {\sigma}^2(t)$.
Thus the distribution gets flatter in time,
suggesting that the differences of wealth may only
grow with time:
the spread between lower and upper end of middle class
increases. This is what one  very often observes if one
surveys a macro-system over years, but not always. There are two
reasons for this. Firstly, the simple model (\ref{mult})
seems to be inappropriate to describe the wealth evolution
in turbulent periods like wars or crises. Secondly,
the mean-field approximation (\ref{mult}) fails
to reflect the conservation law for the total wealth
in the macro-system.
If one assumes that the total wealth $W$ changes much
slower in time than the wealths of individuals
then in a short period one can treat the total
wealth as constant in comparison with the wealths
of individual $w_i$'s. This means however that
$w_i$'s cannot fluctuate
independently of each other as is assumed in the equation (\ref{mult})
because it would violate the conservation law
\begin{equation}
W = w_1 + w_2 + \dots + w_N
\label{ww}
\end{equation}
which tells us that, unless the economy as a whole
produces a new wealth, fluctuations of $w_i$ are not
independent \cite{bb1}. This effect does not allow  fluctuations
of a typical order to grow as fast as the equation
(\ref{ln}) would suggest. Later we shall discuss other
consequences of the presence of the conservation law.

There is another economical factor which one should take
into account when considering the process of wealth
fluctuations (\ref{mult}). In each macro-economy there
is some threshold wealth which one has to posses
to function in the system to fulfill minimal needs.
In welfare economies it is provided by the social security system.
Generally for each macro-economical
system one can assume the existence of a
positive cut-off $w_*>0$ for the minimal wealth
of each individual. It is easy to work out
consequences of imposing the cut-off \cite{ls}
\begin{equation}
w>w_*
\label{cut-off}
\end{equation}
on the multiplicative process (\ref{mult}).
The right-hand side of the equation for the
return is also given by the sum of independent
increments as in (\ref{add}). What changes
is the boundary condition: in the presence of a
cut-off, $R_T$ cannot be smaller than a certain
value $R_*$. One can think of the equation~(\ref{add})
as of a random walk, which in the case of a cut-off
has the lower barrier $R_*$.
Microscopically the model with the barrier and
without the barrier are identical. Thus one can check
that both cases are described by an identical differential
equation but with a different boundary condition.
The equation reads
\begin{equation}
\frac{\partial P_T(R_T)}{\partial T} =
-\bar{r} \frac{\partial P_T(R_T)}{\partial R_T} +
{\sigma}^2 \frac{\partial^2 P_T(R_T)}{\partial R^2_T}\,.
\label{FP}
\end{equation}
By inspection one can check that indeed the probability
distribution $P_T(R_T)$ (\ref{prob}) is a solution
of the equation. In physics, the corresponding equation
is called the Fokker--Planck equation. It describes
a random walk with a drift. The two constants $\bar{r}$
and ${\sigma}^2$ in the equation correspond to the drift velocity
and the diffusion constant and are related to
the mean $\bar{r}$ and the variance ${\sigma}^2$ of
the underlying distribution (\ref{mv}). In the presence of
the cut-off in the boundary condition:  $R_T > R_*$.
the Fokker--Planck equation~(\ref{FP})
possesses a stationary solution $P_T(R)=P(R)$
\begin{equation}
\frac{\partial P(R)}{\partial T} = 0
\end{equation}
if $\bar{r} < 0 $.
The equation obtained by comparing the right-hand
side of (\ref{FP}) to zero can be solved with the
normalization condition
\begin{equation}
\int\limits_{R_*}^\infty P(R) d R = 1\,.
\label{norm}
\end{equation}
The solution reads
\begin{equation}
P(R)  =
\alpha \exp -\alpha (R-R_*)\,,
\label{rt}
\end{equation}
where $\alpha = -\bar{r}/{\sigma}^2 > 0$.
Substituting the return $\bar{r}$ by $w = w_0 e^{R}$
(\ref{add}) one eventually obtains the stationary distribution
for wealth \cite{ls}
\begin{equation}
p(w) d w =
\frac{\alpha w_*^\alpha}{w^\alpha} \frac{d w}{w}\,.
\label{wt}
\end{equation}
Notice that it is independent of $w_0$ which disappears from
the solution. This is the Pareto law \cite{p}. When the drift $\bar{r}$
is positive the exponent $\alpha$ is negative,
the  normalization condition
(\ref{norm}) cannot be fulfilled.
There is no stationary solution.
For positive $\alpha$ the distribution
flows with time and approaches the log-normal
law (\ref{ln}) of the Gibrat universality class \cite{g}.
In this case the traces of the lower limit gradually
disappear due to the positive drift which makes the
bulk of the distribution depart from the lower cut-off.
Now imagine that the drift changes slowly in time
taking sometimes positive and sometimes negative values.
In this case the system oscillates between the Gibrat and
Pareto universality classes.
For a finite time of the system evolution it may effectively
lead to a mixed Pareto--Gibrat properties of the distribution,
being in accordance with empirical observations \cite{s}.

What is counter-intuitive in this picture at the first glance
is that the distribution of average returns $\rho_\varepsilon(r)$
generates the Pareto tail in the outcoming distribution
of wealth when the drift $\bar{r}$ is negative.
We see then that power-law tails occur in the wealth
distribution when the system on the average generates
negative returns. Negative returns mean that people loose wealth.
Thus, paradoxically, when most of the people get poorer
some get extremely rich, populating the Pareto tail.
We shall see this effect more transparently below when discussing
a constraint macro-economy.

To summarize this part of the discussion, the theory of
large numbers explains very well the observed
empirical data. Fluctuations in the empirical data may
be large due to the fact that the empirical histograms
are based on single measurements. Fluctuations may be
particularly large in the tail of the distribution
where there are only few counts in the empirical histograms
and where the wealth fluctuations may be large
due to the fat tails (\ref{pf}).

\section{Wealth condensation}

One of the implications of the mean-field approximation (\ref{mult})
is that the total wealth of the
system might fluctuate with the amplitude proportional
to the amplitude of individual changes and the square root
of the number of individuals, or with a higher power if
the fat tail properties become important. In reality the total
wealth of the macro-system alternates slower in time
and does not undergo such fluctuations. Therefore it is
natural to introduce another time scale for changes of the
total wealth than for changes of individual wealths.
This leads to the constraint of the type (\ref{ww}) in which the value
$W$ on the left hand side changes much slower than $w_i$'s
on the right-hand side. This means that the flow of the
wealth between individuals within the system is  much faster
than the process of change of the total wealth.
Thus, if one considers changes of $w_i$'s in a short time
the constraint (\ref{ww}) means that $w_i$'s cannot
be treated as completely independent stochastic variables. In
particular if an individual becomes very rich, amassing
a substantial part of the total
wealth $W$ accumulated in the macro-economical system,
this happens at a price of making others poorer.
It is instructive to analyze consequences
resulting from the constraint. We shall do this
in the following way. In statistical mechanics
of quasi-stationary systems one approximates averages
over time by averages over a statistical ensemble.
We shall use this approach here to represent fluctuations
of the partition of wealth as a sum over all states in
the ensemble of wealth partitions
with the micro-canonical partition function
\begin{equation}
Z(W,N) \ = \ \sum_{\{w_i \ge 0\}} \, \prod_i \, p (w_i) \,
\delta \left( W - \sum_{i = 1}^N w_i \right) \, .
\label{zwn}
\end{equation}
The total wealth  $W$ (\ref{ww}) is distributed
among $N$ individuals. This model is very close in spirit
to the mean-field approximation discussed above since
it assumes almost entire factorization of the probability
into independent probabilities $p(w_i)$ of individuals.
One could, of course, introduce interactions between
different values $w_i$ and $w_j$ but as discussed above
the mean field arguments are good enough to explain empirical
data within the accuracy provided by single observations.
We use here the strategy of not introducing refinements
which are not necessary.
The full factorization is weakly violated by the wealth
conservation. The individual wealths are bounded from
below $w_i > w_*$. For technical reasons it is convenient
to consider integer valued $w_i$'s. From the economical point
of view this means that there exists a minimal indivisable
unit in which one expresses wealth as for example the monetary
unit used in the country. The only thing we shall assume
about the probabilities $p(w)$,
following the previous section,
is that they possess a Pareto tail (\ref{pareto}).
As will become clear, the details concerning the exact shape
of the probability distribution are irrelevant for the universal
large scale effects of wealth condensation. The only important
parameters of the model are the value of the Pareto
exponent $\alpha$ and the mean of the distribution
\begin{equation}
w_{cr} = \sum w p(w)\,.
\end{equation}
The mean is finite for $\alpha>1$ and infinite otherwise.
In a thermalized economy where $p(w)$ is constant for a long time
this average $w_{\rm cr}$ adjusts itself
to the average {\it per capita}
\begin{equation}
\bar{w} = \frac{W}{N} \,,
\end{equation}
and one has
\begin{equation}
w_{\rm cr} = \bar{w}\,.
\end{equation}
The mean of the distribution $w_{\rm cr}$ may however depart from $w$
as a result of some changes which the system may undergo.
For example it may happen
that for some reasons a thermalized stable economy will
start to develop, increasing the total wealth $W$.
Alternatively the economy may quickly go down decreasing
the total wealth $W$.
The question arises how the system
adjusts to the new situation in which $\bar{w} \ne w_{\rm cr}$:
how it redistributes the surplus if $\bar{w} > w_{\rm cr}$
or covers the deficit if $\bar{w} < w_{\rm cr}$.
A potential discrepancy between $w_{\rm cr}$ and $\bar{w}$ may also
occur as a result of some structural changes of the macro-economical
framework, like taxation laws, employee rights {\it etc.},
which may lead to a change of the distribution $p(w)$ yet before
the total wealth of the economy changes.

We shall try to answer this question by investigating the response
of the system defined by (\ref{zwn}). This model can be solved
analytically \cite{bb0,bb1}. The response of the system
can be determined from the shape of the effective probability
distribution defined as an average over all partitions weighted
 by the partition function (\ref{zwn})
\begin{eqnarray}
\widehat{p}(w) =
\frac{1}{N}\left\langle \sum_i^N \delta(w_i-w)\right\rangle .
\end{eqnarray}
One can show that when $w_{\rm cr} = w_*$, there is a
perfect matching and the effective probability
\begin{equation}
\widehat{p}(w) = p(w)\,.
\end{equation}
However, when
the wealth {\it per capita} exceeds the critical value
$\bar{w} > w_{\rm cr}$ or is smaller than the critical value:
$\bar{w} < w_{\rm cr}$ the system enters one of two different
phases which we call the surplus phase or the deficit phase respectively.

In the surplus phase the effective probability distribution $\widehat{p}(w)$
nonuniformly approaches  $p(w)$ creating a peak at the large values.
For large systems $N\rightarrow \infty$ the effective probability
density may be approximated by
\begin{equation}
\widehat{p}(w) = p(w) + \frac{1}{N} \delta(w - N \Delta w)\,,
\end{equation}
where the second term is the Dirac delta localized at the value
proportional to the system size $N$. The proportionality coefficient
$\Delta w = \bar{w} - w_{\rm cr}$ is a deviation of the average wealth
from the critical value. The coefficient $1/N$ in front of the delta
function means that the probability
related to the peak is $1/N$, or equivalently that the contribution comes
from one out of $N$ individuals.
The wealth of this individual $w_{\rm max} = N \Delta w$ grows with the
system size. He or she takes a finite fraction of the whole wealth.
This effect is similar to the Bose--Einstein condensation for which
a finite fraction of all particles is in the ground state. The difference
between the two condensations is that in the Bose--Einstein condensation
the ground state is favored by the energy,
while here all individuals are identical
and therefore they have {\it a priori} the same chance that the wealth will condense
in their pocket. The condensation results from a spontaneous symmetry breaking
mechanism which breaks the permutation symmetry of $N$ individuals of the
original model. In reality, of course, the position of individuals
in the macro-system is not identical. This may further
enhance the effect of condensation observed already in the model
where those differences are neglected.

In the deficit phase ($ \bar{w} < w_{\rm cr}$)
the effective probability distribution $\widehat{p}(w)$
is given by
\begin{equation}
\widehat{p}(w) = c e^{-\mu w} p(w)\,,
\end{equation}
where $\mu$ is some positive function which depends on
$\Delta w = \bar{w} - w_{\rm cr}$.
The factor $c$ is a normalization constant.
The exponent $\mu$ vanishes in the limit $\Delta w \rightarrow 0^{-}$.
We see that when the system enters the deficit phase
a suppression of the fat tails occurs: these are the richest who
first  pay for the deficit.

The order of the transition between the deficit and surplus phases
depends on $\alpha$. The transition
is of the third or higher order \cite{bb0}.
The transition becomes weaker
when $\alpha$ approaches one or infinity.
The critical value $w_{\rm cr}$ being the average of the distribution
depends on the whole distribution but it is very sensitive to
the tails: the fatter the tail the larger the critical value
$w_{\rm cr}$. On the other hand, when the critical
value $w_{\rm cr}$ is larger it is more difficult to
enter the surplus phase $\bar{w} > w_{\rm cr}$ because the wealth
{\it per capita} must exceed this critical value.
This may happen in a very rich society. In the limiting case
$\alpha=1$, the critical value $w_{\rm cr}$ is infinite and the system
never enters the surplus phase.

When the critical value $w_{\rm cr}$ becomes smaller it is easier for
the wealth {\it per capita} $\bar{w}$ to exceed $w_{\rm cr}$ and to enter the
surplus phase where the system has problems to redistribute the
wealth of the richest. If it happens in a rich society
this means that one individual creates a large fortune and
the system is not able to redistribute it quickly or at least
that such a redistribution is not favored statistically.
The wealth condensation becomes however natural then. It is not a shame
to be rich in a rich society as says Confucius.

Paradoxically, the condensation may also take place
in a restrictive macro-economy. Assume that the total
wealth of a poor society is fixed.
Additionally imagine that the system becomes more restrictive,
which results in the increase of the Pareto index and the
decrease of the critical value $w_{\rm cr}$. If this value
becomes smaller than the wealth {\it per capita} $\bar{w}$, which is fixed,
the system enters the surplus phase. The wealth condensates
in one pocket as a result of the surplus anomaly.
Some of the richest become richer and other poorer. This clearly reveals
the danger of corruption of restrictive poor macro-economies.

The main conclusion of this section is that large number theory
also on the elementary level explains potential danger of
statistical instability, which in the case of restrictive
macro-economy may be related to the phenomenon of corruption.
One can avoid this danger by making the macro-economical rules
more liberal \cite{bb1,bb2}. For completeness
let us  mention that one can consider a macro-economy
in contact with the external world \cite{bb2}.
In the language of statistical physics this
corresponds to the model defined by the canonical
version of the partition function (\ref{zwn}). In addition
to what we discussed here, in the canonical
version of the model one can observe statistical effects of the
attraction of the external wealth to the macro-economy,
or the withdrawal of the internal one,
depending on whether the macro-economical rules inside
or outside are more liberal.

\section{Modeling a financial market}

Let us now turn to the mesoscopic scale and discuss financial markets.
Financial market is a part of the econosystem which is easiest to quantify.
We shall use a simplified picture of this market in which the only objects are
the prices of assets, asset being the name commonly used to describe a
financial instrument, which can be bought or sold, like currencies, bonds,
shares {\it etc.} In the following we shall understand assets solely as shares.
Asset (or stock) prices $S_i(t)$ are functions of time. A typical time step
$\varepsilon$, when the price is changed is as short as few seconds.
It will be the dynamics of price changes, which we shall discuss in this
chapter.

In the analogous way as the quantity $r_t$  (\ref{return}) of the chapter
about macro-economy we define the instantaneous returns, which we
shall alternatively call relative price changes of the asset
in the period from $\tau$ to $\tau+\varepsilon$
\begin{eqnarray}
x_i(\tau;\varepsilon) = \log S_i(\tau+\varepsilon) - \log S_i(\tau).
\label{multagain}
\end{eqnarray}
Again the crucial ingredient of this analysis is the
assumption about the multiplicative nature of price changes.
The definition of return is independent of the unit
in which the price is given and seems the best to capture
the essential properties of the price system.
Return $x_i(\tau;\varepsilon)$ can be any positive real number. Obviously
the return over a larger time interval is
a sum of all changes over its subintervals
\begin{eqnarray}
x_i(\tau;\varepsilon_1+\varepsilon_2)=x_i(t;\varepsilon_1)+x_i(t+\varepsilon_1;\varepsilon_2)\,.
\end{eqnarray}
Financial databases contain huge number of time series of asset prices, sampled
at various frequencies. Phenomenologically one can observe that prices behave
in a random way: relative price changes $x_i(t,\varepsilon)$ fluctuate.
The empirically measured
time correlations show that these fluctuations have a rather short
autocorrelation time, typically of the order of several minutes. Longer
autocorrelation times were observed for the absolute values of fluctuations.

If the frequency of sampling $\varepsilon$ is chosen larger than the
autocorrelation time $\varepsilon_0$,
corresponding price changes can be viewed as independent random variables.
The simplest assumption one can make is the assumption of stationarity:
$x_{it} = x_i(\tau=t*\varepsilon_0;\varepsilon_0)$, where $t$ is an integer,
can be interpreted as random numbers generated with the same random
number generator, independent of time. One can
derive surprisingly strong predictions based on this simple assumption, using
very general properties of this random number generator. Let us assume that
the generator is characterized by the normalized probability distribution
function
(pdf) $P(x)$, with a characteristic function $\hat{P}(z)$ defined by
the Fourier transform
\be
\hat{P}(z) = \int\limits_{-\infty}^{\infty} dx \, P(x) e^{i x z}\,.
\label{ft}
\ee
Define a function $\hat{R}(z) = \log \hat{P}(z)$.
It is straightforward to see that the sum
\be
X_n = \sum_{i=1}^n x_i
\ee
of independent random numbers distributed with $P$
is again a random number with a distribution
$P_n$ being an $n$-fold convolution of $P(x)$. In consequence,
$\hat{P}_n(z) = \hat{P}^n(z)$ and $\hat{R}_n(z)= n\hat{R}(z)$ where
$\hat{R}_n(z) = \log P_n(z)$.

A special role is played by stable distributions, which have the
property that the probability distribution of the sum $P_n$ can
be mapped into the original distribution by a linear change of the
argument
\be
d x \, P_n(x) = d (a_n x + b_n) \  P(a_n x + b_n)\,,
\label{convolution}
\ee
where $a_n$ and $b_n$ are suitable parameters.
Saying differently,
the stable distributions are self-similar under the convolution
which means that the shape of pdf is preserved up to a scale factor
and shift. The condition (\ref{convolution}) can be
rewritten as a condition for $\hat{R}(z)$ in the form
\be
\hat{R}(z) = n \hat{R}(a_n z) + i b_n z\,.
\label{conv1}
\ee
A class of stable distributions is limited. The best known is the Gaussian
distribution, for which
\be
\hat{R}(z) = -\gamma z^2 + i\delta z\,,
\ee
where $\delta = \leftarrow x \rightarrow$ 
and $\gamma = \frac{1}{2} \la (x-\delta)^2\ra$.
One can think of the straightforward generalizations of the last formula
\be
\hat{R}(z) = -\gamma |z|^\alpha + i\delta z\,.
\ee
One can check that they indeed fulfill the stability condition
(\ref{conv1}). However only for $0 <\alpha \le 2$ the corresponding
characteristic function $\hat{P}(z) = \exp \hat{R}(z)$ leads after
inverting the Fourier transform (\ref{ft}) to a positive definite and
normalizable function $P(x)$, which only in this case can be interpreted
as  a probability distribution.

It is a special case of Lévy distributions
characterized by the index $0 < \alpha \le 2$ which can be further
generalized to asymmetric functions.
The most general form of $\hat{R}(z)$ can be shown (\cite{gk}) to be
\be
\hat{R}(z) &=& -\gamma
|z|^{\alpha}(1+i\beta \tan(\frac{\pi\alpha}{2})\rm{sign}(z))+
i\delta z\,,\quad\alpha\ne 1,\nonumber\\ 
R(z) &=& -\gamma |z|(1+i\beta\frac{2}{\pi}\rm{sign}(z)\ln(\gamma|z|) +
i\delta z\,,\ \ \quad\alpha = 1\,.
\label{levy}
\ee
The asymmetry parameter $\beta$ takes values in the range $[-1,1]$.
For $\alpha=2$ we have the Gaussian distribution, the asymmetry
plays no role in this case as one can see from the formula since
the $\beta$-dependent term drops. Indeed the Gaussian distribution has only
a symmetric realization.

One can easily check that for stable distributions the self-similarity
parameter scales as $a_n = n^{-1/\alpha}$.
Although $\hat{R}(z)$ is given explicitly, only in very few cases
the corresponding pdf $P(x)$ is expressible in terms of simple
analytical expressions.
For $x \to \pm\infty$ and $\alpha < 2$
\be
d x \ P(x) \propto d x \ \frac{A_{\pm}^\alpha}{|x|^{1+\alpha}}
\label{assymetry}
\ee
and the asymmetry parameter
\be
\beta=\frac{A_+^\alpha - A_-^\alpha}{A_+^\alpha + A_-^\alpha}\,.
\ee
This behavior means that Levy distributions are very different from the
Gaussian distribution. For $1 <\alpha<2 $ only the first moment $\la x\ra$
is defined, all higher moments diverge. For $0< \alpha \le 1$ even the first
moment diverges.

The importance of the stable distributions is demonstrated by the central
limit theorem. Suppose we start with an arbitrary distribution $P(x)$, not
necessarily stable. Performing the $n$-fold convolution of this distribution,
in the limit $n \to \infty$ we necessarily end up with one of the stable
distributions described above. Typically if $P(x)$ has the asymptotic behavior
like (\ref{assymetry}) for arbitrary $\alpha > 0$ we shall obtain the
Lévy distribution if $\alpha < 2$ or Gaussian distribution if
$\alpha \ge 2$. As a consequence, if our sampling frequency in the price list
is large, say one day, we may expect to a good approximation
the relative price changes measured with this frequency to be
random numbers obtained from one of the stable distributions.

If the idealized assumption of stationarity holds, we can represent the
history of the financial market as a matrix $x_{it}$, with the times
$t$ measured in intervals of the sampling unit $\varepsilon$,
corresponding to one day. In this way we lose information about
the short time scale fluctuations, but
we may expect that for each $i$ the entries $x_{it}$  will represent
a sequence of random numbers drawn from the same stable distribution.
It is, of course, a crucial question, which stable distribution is realized
in practice.  We may deduce the properties of this distribution studying
a finite sample of $x_{it}$ on a time window $T$,
consisting of many days (say one month).

\section{Gaussian world}

Simplest models assume the distribution to be Gaussian. If this is the case,
it can be characterized by two parameters: the shift
$\delta_i = \la x_i \ra$ and the variance $\sigma_i^2 = 2\gamma_i^2
= \la (x_i - \delta_i)^2\ra$. Both parameters can be easily determined
empirically from the data  on a time window $T$ by the following estimators
\be
\tilde{\delta}_i   &=& \frac{1}{T}\sum_t^T x_{it}\,, \nonumber\\
\tilde{\sigma}_i^2 &=& \frac{1}{T}\sum_t^T \left(x_{it}-
\tilde{\delta}_i\right)^2\,.
\label{estim}
\ee
Obviously these numbers
would be subject to a statistical error due to the finiteness of the
time window. The values of the estimators
converge to the exact values $\tilde{\delta}_i \rightarrow \delta_i$,
$\tilde{\sigma}^2 \rightarrow \sigma^2$ only in the limit $T \to \infty$.
In the Gaussian world the evolution of the price
(or in our case the logarithm of the price) is just a diffusion process
with a drift. Knowledge of the parameters of the Gaussian distribution
describing price changes in one day
can be used to predict the distribution of the relative price changes
on a longer time scales. These will again be given by the Gaussian distribution
(due to its stability), but with rescaled variance and shift.

The market consists of many assets (say $i=1,\dots,N$).
The number of assets in the market is typically a large number
(the well-known Standard and Poor index SP500 quotes prices of 500 companies).
The market reality is more complex than suggested by the model
of independent stationary Gaussian returns discussed above.

The first problem is that the market reality is not stationary.
One cannot expect that the prices will fluctuate according to the
same law over twenty years. In this period many things may happen
which may affect performances of individual companies. One has
to weaken the stationarity assumption and to substitute it by
a sort of quasi-stationarity. In practice this means that the time
window $T$ used in the estimators (\ref{estim}) should be limited
and so should be the future time in which one uses the value of the
estimators. Practitioners \cite{RISCMETRIX}
introduce further improvements to the estimators by weighting past
events with weight, which gradually decreases with time. Here we shall not
discuss this issue further, assuming in what follows a quasi-stationarity.

The second correction which one has to introduce to the model
discussed above is that in reality the prices of individual stocks
are mutually correlated as a result of the existence of the network
of inter-company dependencies.
Indeed even by a purely statistical analysis of the correlation
matrix \cite{GOPI} one can observe and determine
the statistical correlations of price fluctuations of stock prices
of companies from the same industrial sectors. Of course, inter-sector
correlations also exist. Further, the stock
market is not a closed system.
The total capital invested in the market may shift between
the stock market and other investments like for instance the real estate.
This leads to the observed periods of flows of the capital
into the stock market or out of the stock market.
As a result the prices may go up or down, depending on whether the
market attracts are repulses the capital. This is closely related
to the effect known in sociology as herding. The effect of herding
is also clearly seen in the statistical analysis of the
matrix which shows the occurrence of an eigenvalue in the spectrum
of the correlation matrix which is significantly larger than all other.
The corresponding eigenvector is interpreted as a vector of
correlations of changes of individual prices to the main
market tendencies which are often referred to as $\beta$-parameters
after the Capital Asset Pricing Model~\cite{capm}. We shall come back
to this issue later. This discussion shows that a realistic approach
should allow to model the inter-company correlations.

A logical generalization of the Gaussian
model described above is the model of correlated asset fluctuations
 generated from some multidimensional Gaussian distribution.
The probability of generating a vector of returns $x_{it}$, $i=1,\dots,N$
at some time $t$ is
\be
\prod_i d x_i \ P(x_1,x_2,\dots, x_N) \sim
\prod_i d x_i \
\exp -\frac{1}{2} \sum_{ij} (x_i - \delta_i) C^{-1}_{ij} (x_j -
\delta_j)\,.
\label{gener}
\ee
The properties of this generator can be assumed, as discussed before,
to be constant in the period of time for which
the shifts $\delta_i$ and the correlation
matrix $C_{ij}$ are estimated (quasi-stationarity)
\be
\tilde{C}_{ij}= \frac{1}{T} \sum_t^T
\left(x_{it}-\tilde{\delta}_i\right)\left( x_{jt}-
\tilde{\delta}_j\right)\,.
\label{ctilde}
\ee
The correlations may be both positive or negative. Knowledge of the
correlation matrix $C_{ij}$ is crucial in financial engineering,
and in the construction of ``optimal portfolios'' following
the Markowitz recipe \cite{MARKOVITZ}. The main idea in
the construction of ``optimal portfolios'' is to reduce
the risk by diversification. The portfolio is constructed
by dividing the total invested capital into fractions $p_i$ which
are held in different assets: $\sum_i^N p_i = 1$.
The evolution of the return of the portfolio is now given
by the stochastic linearized variable
$X(\vec{p}) = \sum_i^N p_i x_i$,
which produces an instantaneous return
$X(\vec{p})_t = \sum_i^N p_i x_{it}$ at time $t$.
The quintessence of the Markowitz idea is to
minimize the fluctuations of the random
variable $X(\vec{p})$ at a given expected return by
optimally choosing the $p_i$'s. The risk is measured
by the variance of the stochastic variable $X(\vec{p})$
\begin{equation}
{\mit\Sigma}^2 = \sum_{ij} p_i C_{ij} p_j \, .
\label{sigma}
\end{equation}
Clearly, the information encoded in $C_{ij}$ is crucial
for the appropriate choice of $p_i$'s. Intuitively, a
diversification makes only sense when one diversifies
between independent components and one does not gain too much
if one redistributes capital between strongly correlated
assets which make collective moves on the market.

The covariance matrix contains this precious
information about the independent components.
The spectrum of eigenvalues tells us
about the strength of fluctuations of individual
components, and the corresponding eigenvectors about
the participation of different assets in this independent
components.

The fundamental question which arises is how good is the
estimate $\tilde{C}_{ij}$ given by the equation (\ref{gener})
of the underlying covariance matrix (\ref{ctilde}), in particular
how good is the risk estimate
\begin{equation}
\tilde{{\mit\Sigma}}^2 = \sum_{ij} \tilde{p}_i \tilde{C}_{ij} \tilde{p}_j \,
\label{sigmat}
\end{equation}
of risk (\ref{sigma}).
Although the question looks simple, the answer
is not immediate. One can quantify the answer with the
help of the random matrix theory. We shall sketch some ideas
which one uses in this theory in the next sections. Here
we shall only quote the results.

To start with, consider the simplest case of completely
uncorrelated assets which are equally risky. Further,
we assume that they all fluctuate symmetrically around
zero $\delta_i=0$ with the same
variance $\sigma_i=1$. The correlation matrix reads in this case
$C_{ij} = \delta_{ij}$. The spectrum of eigenvalues
of this matrix is $\rho(\lambda) = \delta(\lambda-1)$ which means that it
is entirely localized at unity. For the ideal diversification
$p_i = 1/N$ the risk measured by ${\mit\Sigma}$ (\ref{sigma})
is ${\mit\Sigma} = 1/\sqrt{N}$. What shall we obtain if we
use in this case the estimate $\tilde{C}_{ij}$ instead?

The random matrix theory as we shall see later gives a definite
answer. The first observation is that the quality of the
estimator (\ref{ctilde}) depends on the time $T$ for which
we could measure the correlation matrix. The longer time $T$,
the better quality of the information which can be read of from
$\tilde{C}_{ij}$: all diagonal elements should approach unity,
and off-diagonal ones zero. In reality, as we mentioned, one
never has an infinite time $T$ at ones disposal. Geometry of the
data matrix $x_{it}, i=1,\dots,N, t=1,\dots, T$ is finite. It is
just a rectangular matrix with the asymmetry parameter $a=N/T < 1$.
Such matrices form an ensemble called the Wishart ensemble~\cite{WISHART}.
The case $a > 1$ requires a special treatment and is not relevant
in this case.
For $a$ larger than zero we expect that the spectrum of the
matrix $\tilde{C}$ will be smeared in comparison with the
delta spectrum of ${C}$. Indeed, as we shall see
in the next sections using the methods of
random matrix theory one finds
\be
\tilde{\rho}(\lambda) = \frac{1}{2\pi a}
\frac{\sqrt{(\lambda_+-\lambda)(\lambda-\lambda_-)}}{\lambda}
\label{rhot}
\ee
with $\lambda_\pm = (1\pm \sqrt{a})^2$.
Only in the limit $a\to 0$ we get the spectrum peaked at unity.
This spectrum is calculated from the random matrix theory for
Wishart matrices as we discuss later.

Although the empirical matrix $x_{it}$ is obtained
from a single realization of a random matrix
from the Wishart ensemble, its spectral properties are
in general very similar to those described above.
This is due to the self-averaging property of large matrices.

We can also explicitly find the estimate of risk (\ref{sigmat}).
In doing this one should take into account that the optimal
choice of probabilities $\tilde{p}_i$ which minimizes the
risk $\tilde{{\mit\Sigma}}$ depends on $\tilde{C}_{ij}$
\begin{equation}
\tilde{p}_i = \frac{\sum_j^N \tilde{C}_{ij}^{-1}}
                   {\sum_{jk}^N \tilde{C}_{jk}^{-1}}\,.
\end{equation}
Inserting this solution into the formula (44)
we can calculate the minimal value of the estimated risk
\begin{equation}
{{\mit\Sigma}}^2 = \frac{1}{N}
\frac{\int\limits d \lambda \ \rho(\lambda) \lambda^{-2}}
  {\left(\int\limits d \lambda \ \rho(\lambda) \lambda^{-1}\right)^2 }
\end{equation}
which eventually gives
\begin{equation}
{{\mit\Sigma}} = \frac{1}{\sqrt{N}} \frac{1}{\sqrt{1-a}}\,.
\end{equation}

The exact relation between the spectrum of $C_{ij}$ and $\tilde{C}_{ij}$
can be obtained in the limit $N,T \to \infty$, $a=N/T$ fixed.
Again we skip here the derivation and quote only the result.
A simple formula can be obtained for the Green's function
\be
\tilde{\cal G}(z)
= \frac{1}{N}\left< {\rm Tr}\frac{1}{z-\tilde{C}^{-1}} \right>_{\rm W}
\label{Green1}
\ee
which relates it to its counterpart, in the $T\to \infty$ limit:
\be
{\cal G}(t) = \frac{1}{N}{\rm Tr}\frac{1}{t-C^{-1}}\,.
\label{Green2}
\ee
The subscript W means the average over the Wishart ensemble (\ref{gener}).
One finds~\cite{UNPUBLISHED1}
\be
z \tilde{\cal G}(z) = t {\cal G}(t)\,,
\label{Pastur1}
\ee
here $z$ and $t$ are related to each other as:
\be
z &=& t(1-a+at{\cal G}(t))\,.\label{relt}
\ee
These two relations are in fact a concise way to write infinitely
many relations between the moments of matrices $C_{ij}$ and $\tilde{C}_{ij}$.
Let
\be
c_k &=& \frac{1}{N} {\rm Tr} C^{-k},\\ \nonumber
\tilde{c}_k &=& \frac{1}{N} \la {\rm Tr} \tilde{C}^{-k}\ra_{\rm W}.
\ee
On finds
\be
\tilde{c}_1 &=& c_1\,, \nonumber\\
\tilde{c}_2 &=& c_2 + a c_1^2\,, \nonumber\\
\tilde{c}_3 &=& c_3 + 3 a c_1 c_2 + a^2 c_1^3\nonumber\\
\cdots
\ee

At the end of this section let us come to the problem of the large
eigenvalues observed in the spectra of eigenvalues of the
financial covariance matrices $\tilde{C}_{ij}$. The spectra
consist typically of the random part (\ref{rhot}) which is universal
as discussed above and  few large eigenvalues. Among them one is
particularly large. Its value is roughly speaking proportional to
the number $N$ of the assets in the market. The corresponding eigenvector
contains the contribution from almost all $N$ companies on the market.
This eigenvector is called the ``market''. One can relatively easily understand
the source of the appearance of the market in the spectrum in terms
of the herding phenomena which we shortly signaled before.
Imagine that there is a collective behavior of investors on the market
which can be driven by some sociological factors. Mathematically
such a collective movement may be in the simplest version modeled
by the coupling of the individual prices to some common background, for example
by substituting the generator of the vector of prices (\ref{gener}) by
a new generator of the form
\begin{eqnarray}
\prod_i d x_i \ P(\vec{x}) \sim
\prod_i d x_i \
\exp -\frac{1}{2}
\sum_{ij} (x_i - \beta_i m_t) C^{-1}_{ij} (x_j - \beta_j m_t)\,,
\label{hgener}
\end{eqnarray}
where $\beta_i$'s are some constants, and $m_t$ is a common
random variable describing the market movements. This is the basic
idea underlying the CAPM model \cite{capm} mentioned above. One can
check that the largest eigenvalue disappears from the spectrum
leaving the remaining part intact if at each $t$ one subtracts from
each return the market background represented
as the instantaneous average over all companies.

The other large eigenvalues can be attributed to the real strong
correlations between companies. The analysis of the eigenvectors
allows to divide the market into highly correlated clusters,
usually corresponding to companies from the same industrial
sector. For example, one can see that the gold companies form
a cluster which is anticorrelated to the market.

\begin{figure}[htb]
\centerline{%
\epsfig{file=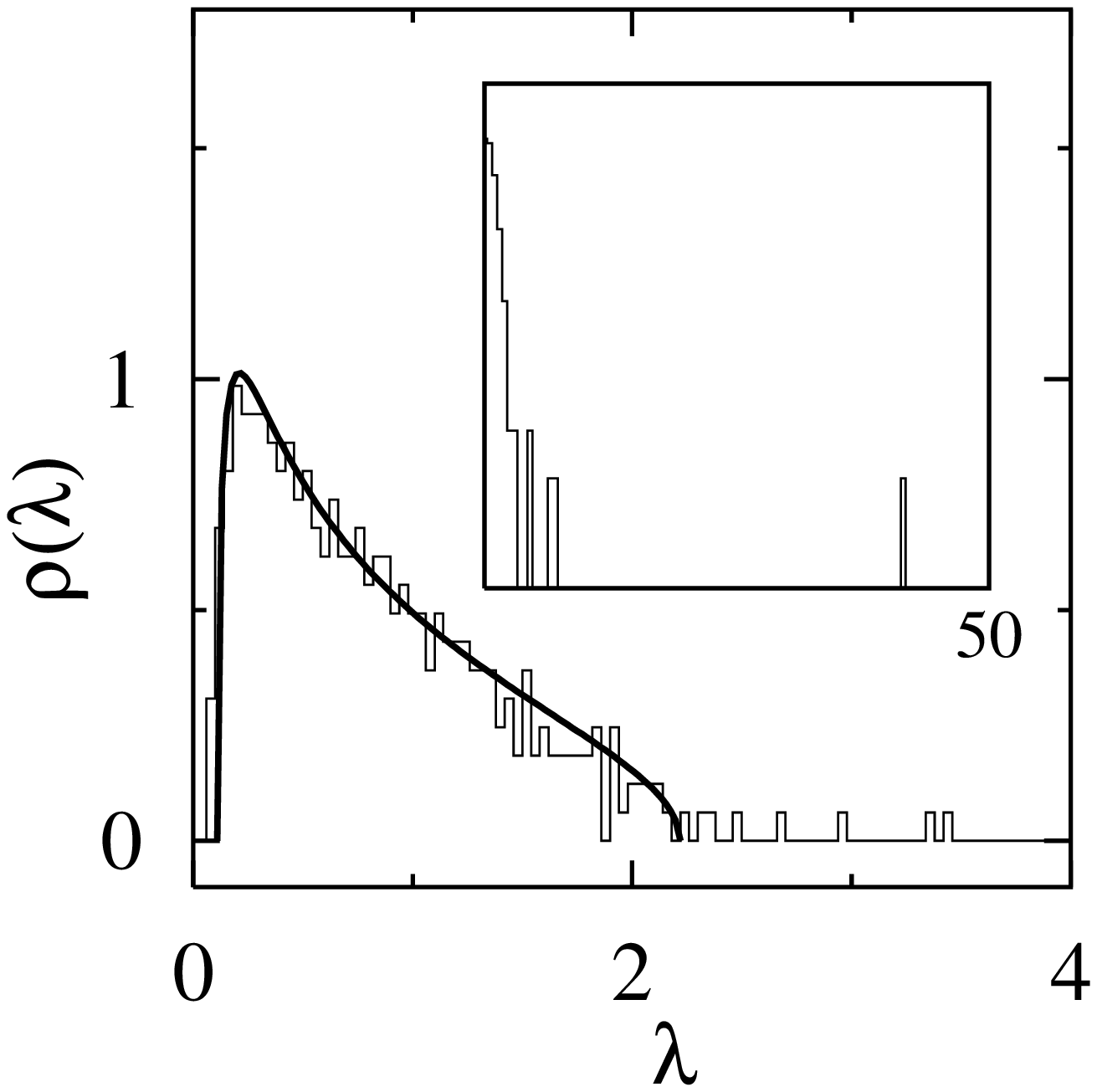,width=7.5cm}
\epsfig{file=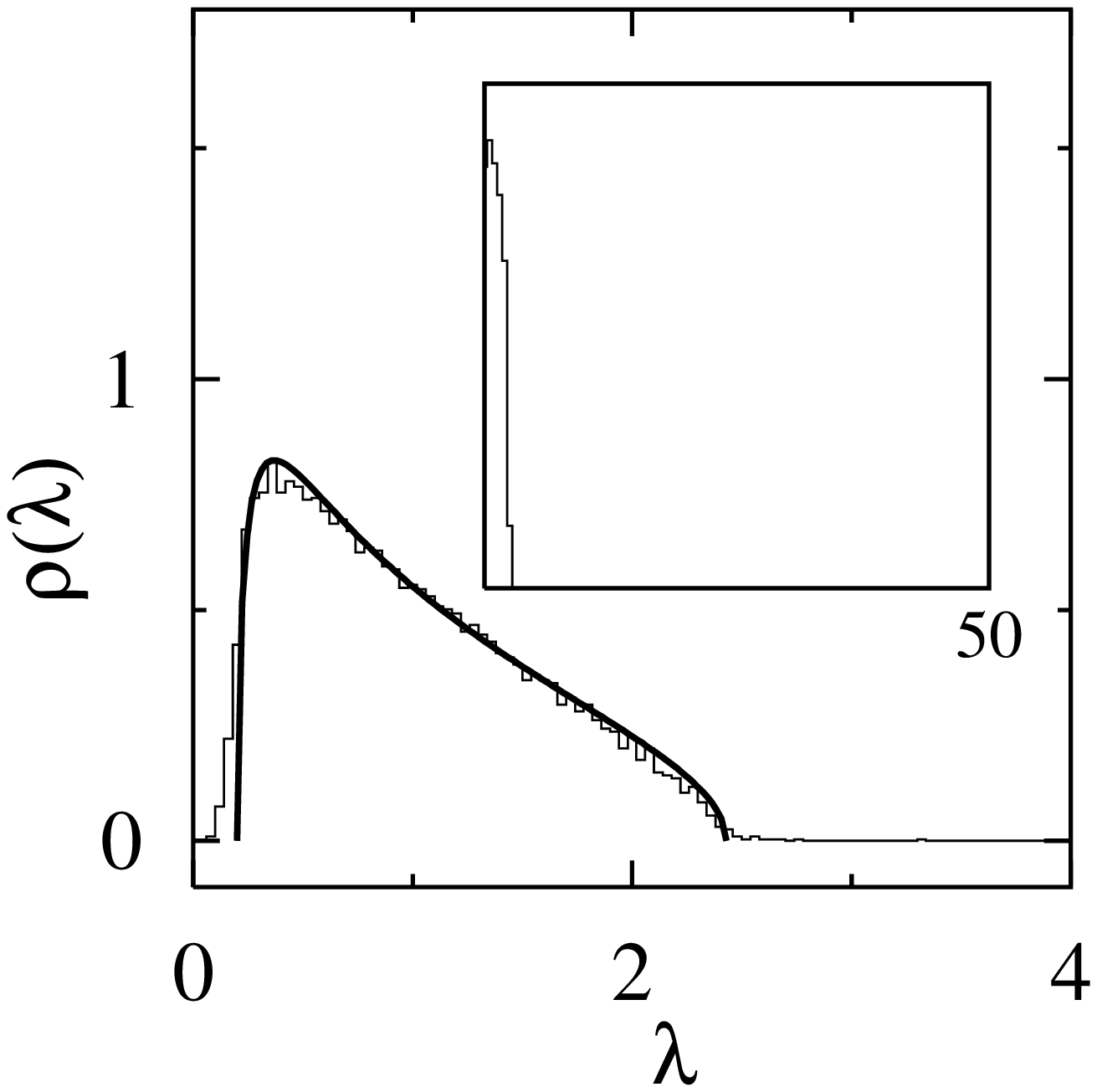,width=7.5cm}}
\caption{The spectrum of the financial covariance matrix
for the daily SP500 for $N=406$ stocks and for
$T=1309$ days from 01.01.1991 to 06.03.1996. The left plot
represents the spectrum of the covariance matrix
for the normalized returns in the natural time ordering;
the right one for the normalized return in the reshuffled
ordering. The reshuffling destroys correlations between entries
of the matrix $\tilde{C}_{ij}$. The random matrix prediction
is plotted in solid line. The
large eigenvalues lying outside the random matrix spectrum
in the left figure disappear from the spectrum for
reshuffled data shown in the right.}
\label{gf}
\end{figure}

\begin{figure}[htb]
\centerline{%
\epsfig{file=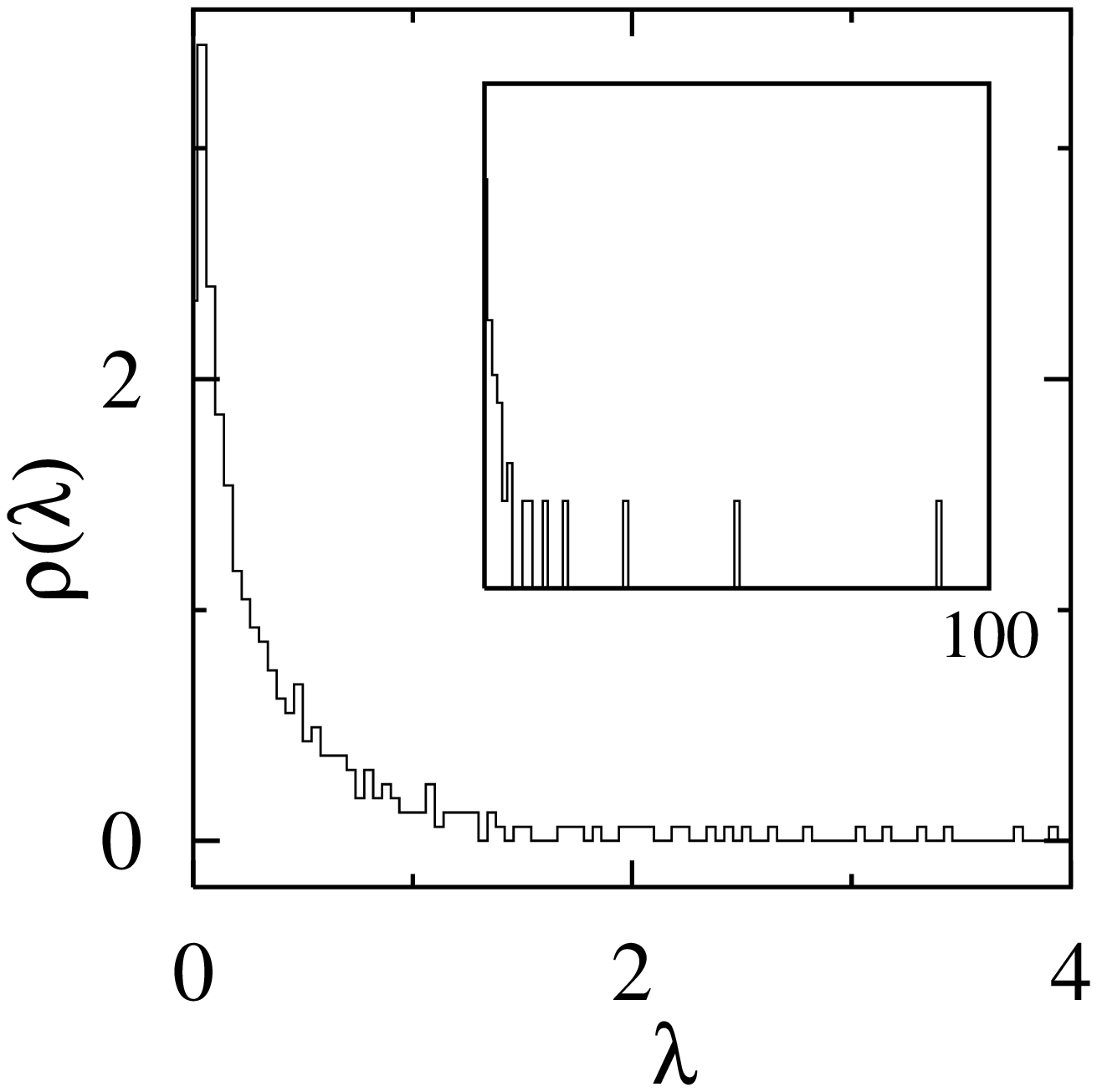,width=7.5cm}
\epsfig{file=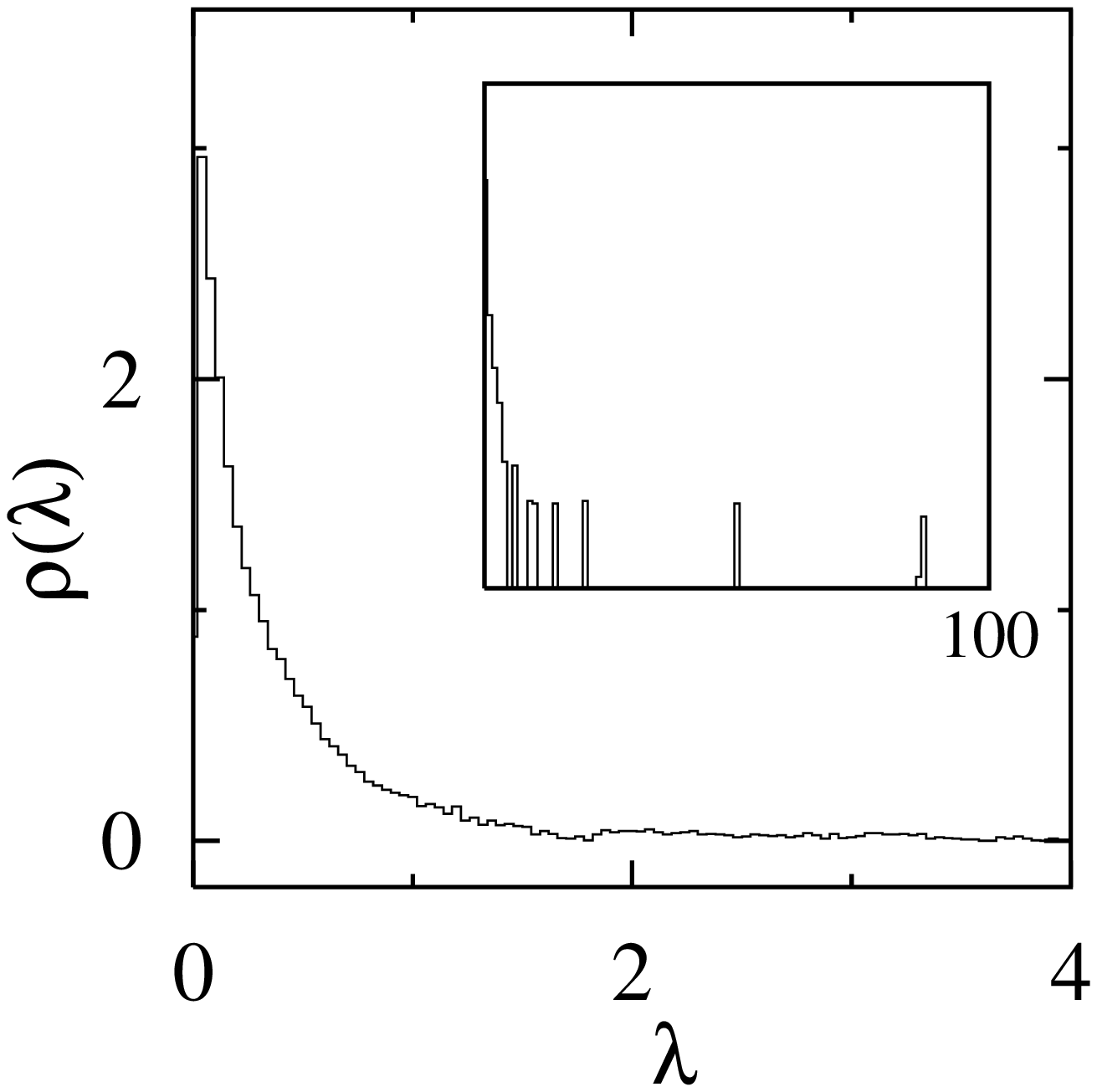,width=7.5cm}}
\caption{The same as in Fig. \ref{gf} but for
for the nonnormalized returns: the left figure for the data
in the natural time ordering and
the right for the reshuffled ordering.
In this case reshuffling does not remove the
large eigenvalues from the spectrum signaling the
presence of non-Gaussian effects in the return statistics.}
\label{gif}
\end{figure}

An example of the eigenvalue spectrum of the
empirical covariance matrix $\tilde{C}$ (\ref{ctilde}),
is shown in figure~\ref{gf}. It is calculated for the
SP500 for the period. The data matrix $x_{it}$ has the
size $N=406$ and $T=1308$ which corresponds to the asymmetry
parameter $a=0.31$. In the spectral analysis of the
empirical matrix one usually
unifies the scale of return fluctuations of different assets
by normalizing them by individual variances $\sigma_i$ (\ref{estim}):
$x_{it} \rightarrow x_{it}/\sigma_{i}$ which for each asset
produces fluctuations of unit width.
For such normalized
fluctuations the formula (\ref{rhot}) tells us that
that the random part of the spectrum of the covariance matrix
should be concentrated between $0.20$ and $2.43$.
We clearly see the presence of larger eigenvalues in the
spectrum presented in the left plot in figure \ref{gif}, which
as mentioned, can be attributed to the inter-asset correlations.
However, the large eigenvalues disappear when one removes the
inter-asset correlation. One can do this by random
reshuffling of the time ordering of returns for each individual
asset. A random reshuffling does not
change the content of information stored in each separate row
of data  but it destroys the statistical information about
the correlations between different rows. Indeed as is shown on the right
plot in the figure \ref{gf}, the larger eigenvalues disappear
from the spectrum. The resulting spectrum of the covariance matrix
of such reshuffled data is perfectly described by the
random matrix formula (\ref{estim}).

The above mentioned normalization of return fluctuation
$x_{it} \rightarrow x_{it}/\sigma_{i}$ is natural if
fluctuations belong to the Gaussian universality class. If the
underlying distributions governing the return fluctuations
have fat tails, this normalization is not
appropriate since the variance of the distribution
does not exist. In this case the use of the normalization
$x_{it} \rightarrow x_{it}/\sigma_{i}$ artificially forces
the resulting rescaled quantities to behave as if they belonged
to the Gaussian universality class of distributions
with the unit variance. This introduces a bias to the analysis
in case of non-Gaussian statistics. Indeed, if one skips this
normalization one observes that
covariance matrices for the original SP500 data as well
as for the reshuffled SP500 data both possess large eigenvalues
in the spectra (see Fig.~\ref{gif}). What is the reason that the
reshuffling does not remove them? Is the random matrix prediction
(\ref{rhot}) wrong? The random matrix prediction is not wrong
of course but is valid only for matrices from the Gaussian ensemble.
The removal of the normalization condition revealed the nature
of the randomness of return fluctuations which contain fat tails.
As we shall discuss later, the spectra of Lévy random matrices
contain fat tails which means that even a completely
random matrix may contain large eigenvalues.
The main conclusion of this discussion
is that the large eigenvalues in the spectrum of financial covariances
stem both from inter-asset correlations and from the Lévy statistics
of return fluctuations and therefore a proper statistical
analysis of financial data, in principle of the eigenvalue
content, would require the new Lévy methodology.

\section{L\'evy world}

Indeed on closer inspection one finds that individual
price fluctuations have rather heavy tails.
Empirically one can fit their distribution, at least in the asymptotic
limit, as a power low of the form (\ref{assymetry}) with the power $\alpha
\approx 1.5\dots 1.8$. Following our earlier discussion this means that
one should rather consider stable Lévy distributions when discussing
the distribution of relative price fluctuations, for the sampling frequency
of the order of one day or more.

Models of this type were proposed in the literature. For a single asset $i$
one should in principle determine four parameters (index $\alpha_i$,
asymmetry $\beta_i$, range $\gamma_i$ and  mean $\delta_i$), which
characterize it's distribution $P_{\beta_i\gamma_i\delta_i}^{\alpha_i}(x_i)$.
In practice such
a determination is numerically very difficult, one can assume a value of
$\alpha$ to be some fixed number in the range given above. Similarly one can
assume the asymmetry $\beta_i=0$ (numerically it is very difficult to
distinguish the effect of asymmetry from that of a non-zero $\delta_i$). Even
with these assumptions the determination of the remaining two parameters is
more difficult, because for Lévy distributions the second moment diverges.

A typical time evolution of the logarithm of price will in the Lévy
world be very different than in the Gaussian world. One observes from
time to time very large jumps, called Lévy flights. The practical
consequence is a relatively large probability of extreme events. Since
these events are responsible for possible large losses on  financial
market, the correct determination of the risk cannot be made if their
probability is underestimated. Each investment on a financial market
is risky and investors must know rather accurately the probabilities
of possible gains and losses.

A Lévy market means that we should describe a multidimensional, possibly
correlated, Lévy random number generator. A natural assumption, as
explained above is a common value of the index $\alpha$ for all market
components. Correlations mean that for a given moment $t_j$, fluctuations
$x_{it}$ can be decomposed as  linear combinations of {\it independent}
Lévy components ${\mit\Lambda}_k,~~k=1,N$,
with a factorizable probability distribution
\be
P(\{{\mit\Lambda}_i\}) = \prod_i P_{B_i{\mit\Delta}_i}^{\alpha}({\mit\Lambda}_i).
\ee
and a unit range ${\mit\Gamma}_i=1$. Such a decomposition means that
\be
x_{it} = \sum_k^N A_{ik}{\mit\Lambda}_k
\ee
and that a probability distribution  of this asset is (because Lévy
distributions are stable) parametrized by
\be
\delta_i &=& \sum_k A_{ik}\Delta_k\,, \nonumber\\
\gamma_i &=& \left(\sum_k |A_{ik}|^{\alpha}\right)^{1/\alpha}\,,~~~{\rm and} \nonumber\\
\beta_i  &=& \frac{\sum_k |A_{ik}|^{\alpha}B_k}{\sum_k |A_{ik}|^{\alpha}}\,.
\ee

In the simplest version described above we may take all $B_k=0$ and
in consequence have all $\beta_i=0$. A matrix $X$ with elements $x_{it},
~~i=1,\dots,N,
~~~t=1,\dots,T$ can be viewed as a single realization of the generalized
Wishart random matrix generated
with the Lévy probability distribution. Determination
of the matrix $A_{ij}$ in this case requires new methods, different
than in the Gaussian case
and will be discussed elsewhere \cite{ouranotherpaper}.

One can construct the analogue of the correlation matrix $\tilde{C}_{ij}$ as
\be
\tilde{C}_{ij}=\frac{1}{T^{2/\alpha}}\sum_t^T x_{it}x_{jt}=
\frac{1}{T^{2/\alpha}}(X X^{T})_{ij}
\ee
and discuss its spectral properties when averaged over the ensemble of
Lévy matrices. The dependence on the size of the window $T$ is different
than in the Gaussian case (which corresponds to the limit $\alpha \to 2$).
To understand the reason for that let us consider the uncorrelated Lévy
matrix with $A_{ij}=\delta_{ij}$. The diagonal elements $d_i=\tilde{C}_{ii}$ are
the sums of squares of the random Lévy variables with the index $\alpha$.
It is trivial to realize that such squares are themselves random variables
and that their distribution has a fat tail with the index $\alpha/2$. Following
the arguments of the central limit theorem
given in the preceding sections we expect that if $T$ is large
enough a sum of such variables will be distributed according to the
corresponding Lévy distribution. We may even argue that this distribution
should by completely asymmetric ($\beta=1$), since the squares are all positive.
The factor $T^{-2/\alpha}$ is the correct scaling factor in this case.
Similar arguments can be used to show that the off-diagonal elements
$\tilde{C}_{ij},~i\ne j$ retain the original index $\alpha$ and therefore
in the limit $T \to \infty$ the eigenvalue spectrum of the matrix
$\tilde{C}_{ij}$ is dominated by its diagonal elements.
The shape of this spectrum is given by the Lévy pdf with the index
$\alpha/2$ and $\beta=1$. This pdf has a power-like behavior with
a relatively low power ($\alpha/2 < 1$) and can easily be responsible
for large eigenvalues, which in this version have no dynamical origin.

To assess the importance of the off-diagonal entries on the
spectrum for finite $T$, we use the standard perturbation theory.
For that, we
write
\begin{equation}
\tilde{C}_{ij} =  \left( d_{i} \delta_{ij}
+ T^{-1/\alpha} a_{ij} \right) \ .
\end{equation}
In the zeroth order, the eigenvalues of $\tilde{C}_{ij}$ are just $d_i$.
The first order corrections are zero because
the matrix $a_{ij}$ is off-diagonal.
Generically, for a random matrix, $d_i$'s are not degenerate,
so up to the  second order, the eigenvalues of $\tilde{C}_{ij}$ are
\begin{equation}
\lambda_i = d_i + \varepsilon^2 \sum_{j (\ne i)} \frac{a^2_{ij}}{d_j-d_i}
= d_i + T^{-2/\alpha} \sum_{j (\ne i)} \frac{a^2_{ij}}{d_j-d_i}\,.
\end{equation}
There are $N-1$ terms in the sum, each of order unity. Thus the
sum contributes a factor proportional to $N$, say $\approx s_i N$,
and we have:
\begin{equation}
\lambda_i =  d_i + s_i N T^{-2/\alpha} \, .
\end{equation}
The off-diagonal terms compete with the diagonal ones for
$N\approx T^{2/\alpha}$.

In the general case, where the matrix $A_{ij}$ is non-trivial, the
usefulness of the correlation matrix $\tilde{C}_{ij}$ to determine the real
correlations in the system is limited. Looking for methods of
determination of the $A_{ij}$ is crucial to distinguish between
the noise and signal.

In both approaches presented above the elements of the matrix $x_{it}$ were
treated as random numbers obtained for each time step $t$ from the same
multidimensional random number generator. This can be understood as a
particular case of a situation where this generator depends also on $t$ and
where we have some non-trivial matrix probability measure $P(x)Dx$. Examples
of such measures are known in the literature.

One can speculate that in reality the distribution of $x_{it}$ comes from
many different sources $s$ and that
\be
x_{it} = \sum_s x_{it}^{(s)},
\label{sumsum}
\ee
where all $x_{it}^{(s)}$ have the same matrix measure. This approach leads
to the concept of non-commutative probability distributions, discussed in
the next chapter.

\section{Matrix economy}

In the previous chapters we mentioned several  consequences
of the central limit theorem,
one of the cornerstones of the theory of probability.
We may ask a question, which at the first glance looks  academic:
Can one formulate an analog of the central limit theorem, if
random variables $\hat{X}_1, \hat{X}_2, \ldots \hat{X}_N$ forming the sums
\be
\hat{S}_N= \hat{X}_1 + \hat{X}_2 +\ldots \hat{X}_N
\ee
{\it do not} commute?
In other words, we are seeking for a theory of probability,
which is non-commutative, {\em ie} $\hat{X}_i$ can be viewed as
operators, but which
should exhibit close similarities to the ``classical'' theory of probability.
Such theories are certainly interesting from the
 point of view of quantum mechanics
or noncommutative field theory, but are they  relevant for economic
analysis?
The answer is positive.
Abstract operators may have matricial representations.
If such construction exists, we would have a natural tool
of formulating the probabilistic analysis directly in the space of matrices.
Contemporary financial markets are characterized by collecting and
processing enormous amount of data.
Statistically, they may come from a processes of the type (\ref{sumsum})
and may obey the matrix central limit theorems.
Matrix-valued probability theory
is then ideally suited for analyzing  the properties of arrays of data (like
the ones encountered in the previous chapter),
analyzing signal to noise ratio and  time evolution of large portfolios.
It allows also to recast standard multivariate
 statistical analysis of covariances~\cite{WILKES}
into novel and powerful language. Spectral properties of large arrays of data
may also provide a rather unique tool for studying chaotic properties,
unraveling correlations and identifying unexpected patterns in very large
sets of data.

The origins of non-commutative probability is linked with abstract
studies of von Neumann algebras done in the 80'. A new twist was given
to the theory, when it was realized, that noncommuting abstract operators,
called {\it } free random variables, can be represented as infinite
matrices~\cite{FRV}.
Only very recently the  concept of FRV started to appear explicitly
 in physics~\cite{GROSS,ZEE,USRANDOM}.

In this paper, we abandon a formal way
 and we shall follow the intuitive approach, using
frequently a  physical intuition.

Our main goal is to study the {\it spectral}
properties of large arrays of data.
Such analysis turned out to be relevant for the source detection
and bearing estimations in many problems related to signal
processing~\cite{SILVERSTEIN}. Since large stochastic matrices
 obey  central limit theorems with respect to their {\it measure},
spectral analysis is a powerful tool for establishing a stochastic
feature of the whole set of matrix-ordered data, simply  by comparing
their spectra to the analytically known results of random matrix theory.
Simultaneously, the deviations of empirical  spectral characteristics
from the spectral correlations  of purely stochastic matrices can be used
as a source of inferring the important correlations, not so visible
when investigated by
other methods.
We shall first formulate the basics of matrix probability theory,
and then we shall discuss a sample application in the case of
a financial covariance  matrix, a key ingredient of any
theory of investment and/or  financial risk management.

Let us assume, that we want to study statistical properties of
infinite random matrices. We are interested in the spectral properties
of $N \times N$ matrix $X$, (in the limit $N \rightarrow \infty$), which is
drawn from a matricial measure
\be
d X \ \exp -N {\rm Tr} V(X)
\label{measure}
\ee
with  a potential $V(X)$ (in general not necessarily polynomial).
We shall restrict ourselves to real symmetric matrices for the moment,
since their spectrum is real.
The average spectral density of the  matrix $X$ is defined as
\be
\rho(\lambda) =\frac{1}{N} \left\la{\rm Tr}\delta(\lambda -X)\right\ra=
\frac{1}{N} \left\la \sum_i \delta(\lambda-\lambda_i)\right\ra\, ,
\label{spect}
\ee
where $\la...\ra$ means averaging over the ensemble (\ref{measure}).
Using the standard folklore, that the spectral properties are related
to the discontinuities of the Green's function
we may introduce
\be
{\cal G}(z) = \frac{1}{N}\left< {\rm Tr}\frac{1}{z-X} \right>\,,
\label{Green}
\ee
where $z$ is a complex variable.
Due to the known  properties  of the distributions
\be
\lim_{\varepsilon \rightarrow 0}
\frac{1}{\lambda \pm i \varepsilon} =PV \frac{1}{\lambda} \mp i\pi \delta(\lambda)
\ee
we see that the imaginary part of the Green's function reconstructs
spectral density (\ref{spect})
\be
-\frac{1}{\pi} \lim_{\varepsilon \rightarrow 0} {\rm Im} \,\,
{\cal G}(z)|_{z=\lambda + i\varepsilon}=
\rho(\lambda)\,.
\ee

The natural (from the point of view of the physicist) Green's function
shall serve us as an auxiliary construction explaining the crucial concepts of the theory
of matrix (noncommutative) probability theory.
Let us  define a functional inverse of the Green's function (sometimes called a
Blue's function~\cite{ZEE}), {\em ie} ${\cal G}[{\cal B}(z)]=z$.
The fundamental object in noncommutative probability theory, so-called
$R$ function or $R$-transform, is defined as
\be
{\cal R}(z)= {\cal B}(z) - \frac{1}{z}\,.
\label{Rf}
\ee
With the help of the $R$-transform we shall now uncover several
astonishing analogies between the classical and matricial
probability theory.

We shall start from the analog of the central limit theorem.
It reads~\cite{FRV}:\\
 The spectral
distributions of independent variables $\hat{X}_i$,
\be
\hat{S}_K = \frac{1}{\sqrt{K}} (\hat{X}_1 +\ldots + \hat{X}_K)
\ee
 each with arbitrary probability
measure with zero mean and finite variance
$\la{\rm Tr } \hat{X}_i^2\ra =\sigma^2$,
converges towards the distribution with $R$-transform ${\cal R}(z)=\sigma^2 z$.

Let us now find the exact form of this limiting distribution.
Since ${\cal R}(z)=\sigma^2 z$, ${\cal B}(z)=\sigma^2 z +1/z$,
so its functional inverse
fulfills
\be
z=\sigma^2 {\cal G}(z) +1/{\cal G}(z)\,.
\ee
The solution of this  quadratic equation (with proper asymptotics
${\cal G}(z)\rightarrow 1/z$ for large $z$)
is
\be
{\cal G}(z)=\frac{z-\sqrt{z^2-4\sigma^2}}{2\sigma^2}
\label{Greengauss}
\ee
so the spectral density, supported by the cut of the square root, is
\be
\rho(\lambda)=\frac{1}{2\pi \sigma^2} \sqrt{4\sigma^2-\lambda^2}\,.
\label{Wigner}
\ee
This is the famous  Wigner semi-circle~\cite{WIGNER}
(actually, semi-ellipse) ensemble.
The omni-presence of this ensemble in various physical applications
finds a natural explanation --- it is a consequence of the central limit
theorem for non-commuting random variables.
Thus the Wigner ensemble is a noncommutative analog of the Gaussian distribution.
Indeed, one can show, that the measure (\ref{measure})
corresponding to Green's function
(\ref{Greengauss}) is $V(X)=\sigma^{-2} X^2$.

Let us look in more detail, what ``independence'' means for two identical
matrix valued ensembles, {\em eg} of the Gaussian type,
with zero mean and unit variance.
We are interested in finding the discontinuities of the Green's function
\be
{\cal G}_{1+2}(z) \sim  \int\limits D\hat{X}_1 D\hat{X}_2
e^{-N {\rm Tr} \hat{X}_1^2} e^{-N {\rm Tr} \hat{X}_2^2}
{\rm Tr } \frac{1}{z-(\hat{X}_1+\hat{X}_2)}\,.
\ee
In principle, this requires a solution of the convolution, with
matrix-valued, noncommuting entries! Here we can see how the $R$-transform
operates. This is the transform, which imposes the  additive property for the
all cumulants: all spectral cumulants  obey $k_i(X_1+X_2) =k_i(X_1) +k_i(X_2)$, for all
$i=1,2,\dots, \infty$~\cite{FRV,SPEICHER}.

Mathematicians call such a property ``freeness'',
hence the name free random variables. The $R$-transform is
an analog of the
logarithm of the characteristic function~(\ref{ft})
in the classical probability theory,
and fulfills the  addition law~\cite{FRV}
\be
{\cal R}_{1+2}(z)={\cal R}_1(z) + {\cal R}_2(z)\,.
\ee
Note that we keep the notation underlying the similarities between
the classical and non-commutative (matricial) probability calculus.
 In the above example, the matrix valued convolution
 of two Gaussian ensembles with a unit
variance gives again a Gaussian ensemble,
with the spectrum (semi-circle) rescaled by $\sqrt{2}$.
Technically, it comes from the fact that ${\cal R}_{1+2}(z)=
{\cal R}_1(z)+ {\cal R}_2(z)= z+z=2z$.
This is like the usual
convolution of two  Gaussian probability distribution, forming also a Gaussian
but with a variance rescaled by a factor $\sqrt{2}$.

At this moment one can start to really appreciate the power of the
noncommutative approach to probability. For large matrices $\hat{X}$ and
$\hat{Y}$
(exact results
hold in the $N=\infty$ limit), the knowledge of their spectra is usually
sufficient for predicting the spectrum of the sum $\hat{X}+\hat{Y}$.

The noncommutative calculus allows also to generalize the additive
law  for non-hermitian matrices~\cite{USNONH,FZEE},
 and even formulate the multiplicative law,
{\em ie} infer the knowledge of all moments of the spectral function
of the product of  $\hat{X} \hat{Y}$,
knowing only the spectra of $\hat{X}$ and $\hat{Y}$ separately
(so-called $S$-transform)~\cite{FRV}. As such,
it offers a powerful shortcut in analyzing
stochastic properties of large ensembles of data. Moreover, the larger the
sets the better, since finite size effects scale at least as $1/N$.

Let us check the possibility of appearance of power-like spectra
in non-commutative probability theory.
Motivated by the construction in classical probability,
we  pose the following
problem: What is the most general form of the spectral distribution
of random matrix ensemble, which is stable under matrix convolution,
{\em ie} has the same functional form as the original distributions, modulo
shift and rescaling?
Surprisingly, non-commutative probability theory
follows from the
Lévy--Khinchine theorem of stability in classical probability.
In general, the needed ${\cal R}(z)$ behaves like $z^{\alpha-1}$, where
$\alpha \in (0,2]$.
More precisely,  the list is exhausted by
the following $R$-transforms~\cite{BERPAT}:
\begin{itemize}
\item[{\it (i)}] ${\cal R}(z)=e^{i \pi \phi} z^{\alpha -1}$, where $\alpha \in (1,2]$, $\phi \in
[\alpha-2,0]$ \\[-2mm]
\item[{\it (ii)}] ${\cal R}(z)=e^{i \pi \phi}z^{\alpha -1}$, where $\alpha \in (0,1)$, $\phi \in
 [1, 1+\alpha]$ \\[-2mm]
\item[{\it (iii)}] ${\cal R}(z)= a+b\log z$, where $b$ is real,
${\rm Im} a \ge 0$ and $b \ge -\frac{1}{\pi} {\rm Im} a$.
\vspace{-2mm}
\end{itemize}

\bigskip

Note that the stability index $\alpha$ is restricted
to  precisely the same values as in the one-dimensional case~(\ref{levy}).
The asymptotic form of the spectra is power-like, {\em ie} $ \rho(\lambda)
\sim 1/\lambda^{\alpha-1}$. Singular case  {\it (iii)} corresponds,
in a symmetric case $(b=0)$, to the Cauchy distribution.
Note that  the case {\it (i)} with $\alpha=2$ corresponds to the Gaussian
ensemble.
For spectral distributions, several other analogies to Lévy
distributions hold.
In particular, there is a one-to-one correspondence for spectral analogs of
ranges, asymmetries and shifts. Spectral distributions exhibit also
duality laws ($\alpha \rightarrow 1/\alpha$),
like their classical counterparts~\cite{BIANE,USLEVY}

To convince the reader, how useful the formalism of non-commutative
probability theory could be for the   analysis of financial data, let us
reconsider the example from the previous chapter.

We analyze a time series of prices of $N$ companies, measured at equal
sequence of $T$ intervals. The returns
(here relative daily changes of prices) could be recast into
$N \times T$ matrix  $X$. This matrix defines the empirical $N \times N$
covariance matrix $\tilde C$ (60).
This matrix  forms today a cornerstone of every methodology of measuring the
market risk~\cite{RISCMETRIX}.

We can now confront the empirical data, assuming the extreme scenario,
that the covariance matrix is completely noisy (no-information),
{\em ie} $X=\hat{X}$ is stochastic, belonging to {\em eg} a random matrix ensemble.
By central limit theorems, we can consider either matricial
Gaussian or matricial Levy--Khinchin
stability basins. From technical point of view, the problem of finding
spectral distribution for covariance matrix reduces
to convolution of a square $T\times T$ matrix $\hat{X}^2$
 and a ``deterministic''
diagonal
projector $P$, with the first $N$ elements equal to 1, and the
remaining $(T-N)$
set to zero. Exact formula, corresponding to $T,N \rightarrow \infty$,
$N/T=a$ fixed comes from a ``back-of the envelope''
calculation~\cite{USWAMB}.
For symmetric Lévy distributions, for completely random matrices, the
Green's function is given by
\be
{\tilde{\cal G}}(z)=1/z[1+f(z)]\,,
\ee
where $f(z)$ is a multivalued solution of a transcendental equation
\be
(1+f)(f+a)\frac{1}{f^{2/\alpha}}=z\,.
\label{cov}
\ee
In the case  $\alpha=2$, equation is algebraic (quadratic), and the spectrum
is localized on a finite interval. In all other cases the range of the
spectrum is infinite, with the large eigenvalue distribution scaling as
$1/\lambda^{\alpha+1}$.
\begin{figure}
\centerline{%
\epsfig{file=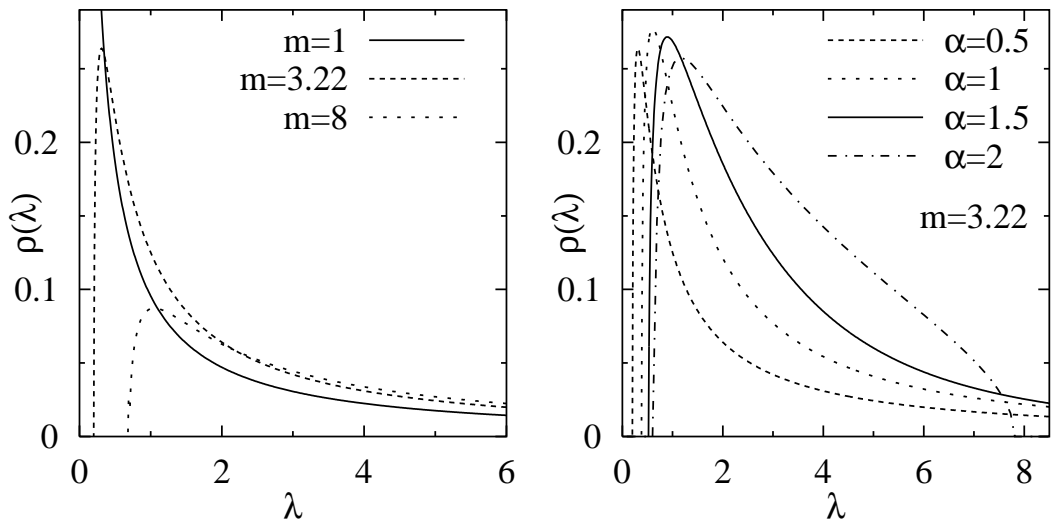,width=14cm}}
\caption{Spectral densities of the covariance matrix
of free random Lévy matrices with
the stability index $\alpha=1/2$ and different
values of the asymmetry parameter by $m=T/N=1/a$
(left figure); and with the given asymmetry parameter
$m=T/N=3.22$ and different values of the stability
index $\alpha$ (right figure).}
\label{frvf}
\end{figure}

A reader familiar with  methods of multivariate statistical analysis
immediately recognizes, that the case $\alpha=2$ corresponds to
the spectral distribution of celebrated Wishart distribution.
Indeed, the normalized solution of a quadratic equation ({\em ie} (\ref{cov})
with $\alpha=2$) leads to  the spectral function (\ref{rhot})
mentioned already.
This  result was rediscovered several times in the context of various
physical applications, with the help of various random
matrix techniques~\cite{WISHARTBIS}.

We would like to stress, how natural and fundamental is this result
from the point of view of non-commutative probability and central
limit theorems.

{}From this point of view, it is also puzzling how late the random matrices
(in our language matricial probabilities) were used for the analysis of
financial data. The breakthrough  came in 1999, when two
groups~\cite{BOUMATRIX,STANMAT} have  analyzed
the spectral characteristics of empirical covariances,
calculated for all companies  belonging to Standard and Poor 500 index,
which remained listed from  1991 till 1996.
 The spectrum of the empirical
covariance matrix constructed from this matrix was then confronted
with the analytically known spectrum of a  covariance matrix
constructed solely from the maximal-entropy (Gaussian) ensemble with
the same
number of rows and columns.

The unexpected (for many) results showed, that the majority of the
spectrum of empirical covariance matrices is populated by noise!

In the case of a Gaussian disorder, 94$\%$ of empirical eigenvalues
were consistent with random matrix spectra~\cite{BOUMATRIX}.
 Only few largest eigenvalues
did not match the pattern, reflecting the appearance
of large clusters of companies, generally corresponding
to the sectorization of the market and market itself~\cite{GOPI}.
The analysis done with the power law ($\alpha=1.5$) not only confirmed
the dominance of stochastic effects, but even interpreted the clusters
as possible large stochastic events~\cite{PRAG}.
 It also pointed at the dangers
of using the covariance matrix (which assumes
 implicitly the finite dispersion) in the case when power laws are present.

The random matrix analysis posed therefore
a fundamental question for quantitative finances.
If empirical covariance matrices are so ``noisy'', why
there are so valuable for practitioners?
Every industrial application of risk measurement depends heavily on
covariance matrix formulation. The
Markowitz's theory of diversification of investment portfolios
depends crucially on the information included in the covariance
matrix~\cite{MARKOVITZ}.
If indeed the lower part of the covariance matrix spectrum has practically
no information, the effects of noise would strongly contaminate the
optimal choice of the diversification, resulting in the
dangerous underestimation of the risk of the portfolio.

Bouchaud and others~\cite{FILTER} suggested a way out, simply filtering out
the noisy part of the correlation matrix and repeating the Markowitz
analysis with refined matrix. This resulted in a better approximation
of the risk.

Their analysis did not answer however the  fundamental question.
If the original matrix is noisy, {\em ie} has almost no information, how come the
covariance matrices form the pillars of quantitative finance?

We tried to answer this question in the previous section,
shedding some light on a rather nontrivial  relation between the true
covariance matrix $C$ and its estimator $\tilde{C}$.
The relation
between the Green's functions ${\cal G}$ and $\tilde{\cal G}$  was
obtained in the framework of Random Matrix Theory.
Some other recent  papers using tools of random matrix theory
for investigating the properties of covariance matrices are
~\cite{IMRE,DROZDZ,GUHR,SORNETTE}.

We would like to point out at this moment, that matrix probability theory
seems to be ideally suited tool for better understanding
the role of covariance matrix and a way of quantitatively
assessing the role of the noise,
important  correlations and the stability of the analysis.
In  our opinion, the full power of random matrix techniques
was not recognized yet by the quantitative finance community.

Finally, we would like to point out an exciting possibility of
introducing the
dynamics formulated in the matrix probability language.
The simplest dynamics of price ($S$) movement of the asset  is
canonically~\cite{OSBORNE}
described by the stochastic equation
\be
\label{sam}
dS= S_{t+dt}-S_t=(\mu dt +\sigma d\eta)S_t\,,
\label{multer}
\ee
where the deterministic evolution is governed by the interest rate (drift)
$\mu$ and the stochastic term is represented by the Wiener measure $d \eta$,
multiplied by dispersion (called in finance volatility) $\sigma$.
The Wiener measure could be realized as $\sqrt{dt} N(0,1)$,
where $N(0,1)$ is a Gaussian with zero mean and unit variance.
Therefore $\langle d\eta\rangle=0$ and $\langle (d \eta)^2\rangle =dt$, reflecting the
random walk character of the process. Since the process is multiplicative,
the resulting Fokker--Planck
equation is
a heat equation with respect to the $\log\, S$, solved by
the log normal
distribution.
Note, that (\ref{multer}) has the {\it same} content as already written
equations  (\ref{mult}),(\ref{multagain}) for wealth and prices, respectively.

One is tempted to write a similar  stochastic equation
for the {\it vector} of prices.
The standard extension~\cite{HULL} reads
\be
S_{t+dt,i}=(1 +\mu_i dt +\sqrt{dt}A_{ij}\eta_j)S_{t,i}\,,
\ee
where the noise {\it vector} $\eta_i$ obeys $\langle\eta_i \eta_j\rangle \sim \delta_{ij}$
and $A_{ij}$ is the square root of the  correlation matrix.

Note however, that one may write a different equation, but now
for the {\it matrix} analog of the Wiener measure.
It is not difficult to see,
that the role of the  white noise is now played by Gaussian ensemble of
random matrices, resulting into the matrix evolution for the whole
{\it vector} of prices. Taking the finite time step, we get
\be
S_{t+dt,i}=(\delta_{ij}+ \mu_{ij} dt  + \sigma \sqrt{dt}
X_{ij})S_{t,j}\,,
\ee
where
$\mu$ is a deterministic matrix and $X$ is a real Gaussian matrix
and not a vector.  Diffusion takes then  place in the space of
matrices.
Finite time evolution results in the infinite product of
large, {\it non-commuting}  matrices,
ordered along the diffusive path, similarly like the chronological
operators do  for the time evolution of non-commuting Hamiltonians.
Here, however, the evolution is dissipative (spectrum is complex).
Surprisingly, random matrix techniques~\cite{IPRO}
allow to analyze the changes
of the spectrum of such stock market evolution operators as a function of time
$t$, similarly as in the case of a single asset, where
the lognormal packet spreads according to the  heat equation.

This approach, basically equivalent
to one of the matrix generalizations of the
Ito-like processes,
 may  allow to study the time properties of the spectra
of large sets of financial data. Moreover, the method
 seems not to  be restricted to the
Gaussian world, due to the mathematical power of matricial probability
calculus and the matrix valued stochastic differential equations
may turn  out  to be a powerful tool of time series analysis of large sets
of data. This ``matrix econophysics'' (as a witticism, or maybe ``wittencism'', we may
use abbreviation M-econophysics to paraphrase M-theory)
may also give a rather precise meaning of
``quantum economy'', a vague term often encounter in the literature.
In the language of a matrix-valued probability calculus,
 the ``quantum nature'' comes from the fact, that basic
objects  of the probability calculus are operators, represented as
large, non-commuting matrices, represented in economy by
arrays of data.  The relevant observables in this language
are related to the statistical properties of their spectra.

\section{Econophysics or econoscience?}

In the course of the presentation, we only briefly analyzed
some selected methods related to the description of real complex systems
such as economic or financial markets.
 The idea was to give the reader not familiar
with this field some sort of a  sampler, hopefully an appetizer.
We did not mention at all several intriguing attempts
to describe financial  crashes using the insight from physics~\cite{CRASHES}.
 Neither did we  mention promising
attempts to use the concepts of cascades and/or
turbulence for explaining the observed
correlations and multifractality in high frequency time series~\cite{BACRY}.
We omitted natural, from the point of view of the
physicist,  modifications of the option theories~\cite{BOUPOT}.
Our presentation of
macroeconomic applications was restricted to simple patterns of wealth
distribution, and we ignored the whole dynamics of this process.
We did not discuss several other issues, usually covered by econophysics
conferences~\cite{DUBLIN,PRAGA}.

At this moment, instead of continuing the list of our sins, let us
come back to the titular question --- how ``solid'' is econophysics as a science?
We would like to point at few dangers, which in our opinion, every
econophysicist has to take into account.
\begin{enumerate}
\item First, we believe that laws of physics do not change in time.
Certainly, this is not true for most of the laws of economy.
Most dramatic are the financial markets.
Technical developments (computers, Internet) or legal regulations
have a major impact
on the field.
\item Second, ``the material points'', {\em ie} agents are not passive ---
they are thinking entities, and sometimes they are very smart.
This invalidates
immediately the ``stationarity'' principle.  Methods and strategies
evolve continuously in time, and the ``quasistationarity'' is rather
due to the traditional conservatism of financial institutions.
Abandoning this conservatism leads to the situation, where more adequate
are concepts of biological evolutionism mixed with elements of the game theory.
Indeed, this lead is seriously studied nowadays~\cite{FARMER,ZHANG}.
Taking into account the complexity of the system, the speed
 at which the systems may evolve and the multidimensional space
of the systems, whose topology may more reflect  the virtual
network of connections than real geographic
distances~\cite{BARABASI,BURDAKRZY},
the need of
 such studies is obvious. As recently pointed~\cite{MIROWSKI2},
economy may  evolve into cyberscience.
Then, the role of the methods of
physics
will be reduced, and physics will serve as a source of
complementary methodology with respects to the methods of
biology, mathematics, psychology and computer science.
\item Even assuming the methods of physics are applicable at certain
   time horizons, econophysics may not be immediately
  successful in the sense of
 making an impact on economic or financial markets.
  What seems to be absolutely crucial is that not only physicists should
   be convinced that they understand ``markets'',
 they have also to convince about that the ``market makers''.
This requires several ingredients.
  The first is the {\it quality} of the research.
  The second is the continuous verification of models/theories with the data.
  The third is the {\it close}
cooperation between the physicists and economists and
 financial advisors.
\end{enumerate}

All these three ingredients are often difficult to fulfill.
  The semantic discrepancies, much too carelessly (also by us)
  usage  of physicists' slang
  (like  quantum economy,
    gauge theory, stock market Hamiltonian,  spin-glass portfolio {\it etc.}), some
   mutual gaps in education, sometimes lack of
  crucial data {\it etc.},  may trigger the situation, where econophysics
   may start to evolve in ``splendid isolation'' from the
 mainstream of economy.

All these dangers may slow down, the however {\it unavoidable} on long run,
 (in our opinion),
impact of methods of physics on economy and financial markets.
Historical  definition of economy, as an {\it art}
 of ``optimal allocation of scarce resources to given ends'',
needs to be replaced by  the {\it science} of
``economic agents --- processors of  information''~\cite{MIROWSKI2}.

We do hope, that this review at least partially convinced the
sceptical reader, that the concepts of statistical physics can
enrich this science, hopefully making even a major impact
at the fundamental level.
\bigskip

The content of this review was greatly influenced by our collaborators,
with whom some of the original work was done and with whom we had
extensive discussions. In particular we would like to thank
Piotr Bialas, Ewa Gudowska-Nowak, Romuald Janik,
Des Johnston, Marek Kamiñski, Andrzej Krzywicki,
Gabor Papp and  Ismail Zahed. We thank Wataru Souma for the
correspondence
and  kind
permission for reprinting the figure from his paper.
This work was supported in part by the  grant
2~P03B~096~22 of the Polish State Committee for Scientific Research (KBN) in years 2002--2004,
EC Information Society Technologies  Programme IST-2001-37259
{\it Computer Physics Interdisciplinary Research and Applications}
and a special dedicated grant of KOPIPOL.

\end{document}